\documentclass[11pt]{article}
\usepackage{graphicx}
\usepackage{epsfig}
\usepackage{amsfonts,amssymb,amsmath} 
\usepackage{latexsym}
\usepackage[english]{babel}

\newcommand{\be}{\begin{equation}}
\newcommand{\ee}{\end{equation}}

\newcommand{\beq}{\begin{eqnarray}}
\newcommand{\eeq}{\end{eqnarray}}

\newcommand{\calP}{{\cal P}}

\newcommand{\calS}{{\cal S}}

\newcommand{\T}{{\cal T}}

\newcommand{\eg}{{\it e.g.,}\ }
\newcommand{\ie}{{\it i.e.,}\ }
\newcommand{\p}{\partial}

\newcommand{\lp}{\left(}
\newcommand{\rp}{\right)}
\newcommand{\lpp}{\left[}
\newcommand{\rpp}{\right]}

\newcommand{\tE}{\widetilde{E}}

\newcommand{\tJ}{\widetilde{J}}
\newcommand{\tS}{\widetilde{S}}
\newcommand{\tT}{\widetilde{T}}
\newcommand{\tO}{\widetilde{\Omega}}

\newcommand{\tloc}{\mathcal{T}}

\newcommand{\ri}{R_\mathrm{i}}
\newcommand{\ro}{R_\mathrm{o}}
\newcommand{\vi}{v_\mathrm{i}}
\newcommand{\vo}{v_\mathrm{o}}
\newcommand{\z}{\tilde{r}}

\newcommand{\pdiffc}[3][\rule{0mm}{0mm}]{\left (\frac{\partial #2}{\partial {#3}}\right )_{\!\!#1}}
\newcommand{\lrp}[1]{\left ( #1 \right )}

\newcommand{\lrpp}[1]{\left [ #1 \right ]}

\begin{document}

\setlength{\unitlength}{1mm}

\thispagestyle{empty}
 \vspace*{0.5cm}

\begin{center}

{\bf \large Bifurcation of Plasma Balls and Black Holes to Lobed
Configurations}

 \vspace*{1.5cm}

{\bf Vitor Cardoso,}$^{1\,,\,2}\,$ {\bf \'Oscar
J.~C.~Dias,}$^{3\,,\,4}\,$

\vspace*{0.5cm}

{\it $^1\,$ CENTRA, Dept. de F\'{\i}sica, Instituto
Superior T\'ecnico, \\
Av. Rovisco Pais 1, 1049-001 Lisboa, Portugal}\\[.3em]
{\it $^2\,$Dept. of Physics and Astronomy, The University of
Mississippi, \\
University, MS
38677-1848, USA}\\[.3em]
{\it $^3\,$DAMTP, Centre for Mathematical Sciences, University of
Cambridge,\\
Wilberforce Road, Cambridge CB3 0WA,
United Kingdom}\\[.3em]
{\it $^4\,$Dept. de F\'{\i}sica e Centro de F\'{\i}sica do Porto,
Faculdade de Ci\^encias da Universidade do Porto, Rua do Campo
Alegre 687, 4169 - 007 Porto, Portugal}\\[.3em]
%
%

\vspace*{0.3cm} {\tt vcardoso@fisica.ist.utl.pt,
O.Dias@damtp.cam.ac.uk}

\vspace*{1cm}

 {\bf ABSTRACT}
\end{center}
At high energy densities any quantum field theory is expected to
have an effective hydrodynamic description. When combined with the
gravity/gauge duality an unified picture emerges, where gravity
itself can have a formal holographic hydrodynamic description. This
provides a powerful tool to study black holes in a hydrodynamic
setup. We study the stability of plasma balls, holographic duals of
Scherck-Schwarz (SS) AdS black holes. We find that rotating plasma
balls are unstable against $m$-lobed perturbations for rotation
rates higher than a critical value. This unstable mode signals a
bifurcation to a new branch of non-axisymmetric stationary solutions
which resemble a ``peanut-like'' rotating plasma. The gravitational
dual of the rotating plasma ball must then be unstable and possibly
decay to a non-axisymmetric long-lived SS AdS black hole. This
instability provides therefore a mechanism that bounds the rotation
of SS black holes. Our results are strictly valid for the SS AdS
gravity theory dual to a SS gauge theory. The latter is particularly
important because it shares common features with QCD, namely it is
non-conformal, non-supersymmetric and has a
confinement/deconfinement phase transition. We focus our analysis in
$3$-dimensional plasmas dual to SS AdS$_5$ black holes, but many of
our results should extend to higher dimensions and to other gauge
theory/gravity dualities with confined/deconfined phases and
admitting a fluid description.

\noindent


\vfill 
\newpage

\tableofcontents

\newpage
\setcounter{equation}{0}
\section{Introduction}
The equations of fluid dynamics and properties of fluids at large
have been used for centuries, not only to describe fluids but also
as {\it analogue} models for other more complex phenomena. For
instance, early experiments with liquid drops by Plateau
\cite{plateau} were aimed at understanding the effect of gravity on
planets (surface tension was then a model for the gravitational
force). Another well-known example, is Bohr and Wheeler's
\cite{bohrwheeler} proposal to describe nuclear fission as the
rupture of a charged liquid drop, where now the surface tension
plays the role of nuclear forces. In general relativity, the
membrane paradigm, whereby a black hole horizon is mimicked by a
stretched fluid membrane, provides another example of the power of
analogue models, with useful applications in astrophysical systems.
Still in a gravity setup, it was recently suggested to use fluid
analogs to explain phenomena observed in general relativistic
scenarios, in particular the classical instability of black strings
and branes \cite{Gregory:1993vy}. Accordingly, the gravitational
Gregory-Laflamme instability would have a counterpart in the
Rayleigh-Plateau instability of fluid mechanics
\cite{Cardoso:2006ks,Cardoso:2006sj} (responsible for the breakup of
liquid jets and tubes).

\subsection{Dual hydrodynamic description of gravity}

The anti-de Sitter/Conformal Field Theory (AdS/CFT) correspondence
adds an interesting twist to this story, making these analogies
powerful and formal. Indeed, it has emerged with the work of Fermi
\cite{Fermi:1950jd}, Landau \cite{Landau:1953gs} and others, that
often the complicated time dependent dynamics of quantum fields is
approximated by a fluid model description. It seems in fact that any
gauge theory has a hydrodynamic limit \cite{landau,HydroPrevious}.
Combining these ideas together one expects gravity to have a {\it
dual} hydrodynamic description \cite{HydroPrevious}. This
expectation has been formally verified in
\cite{Bhattacharyya:2008jc,Bhattacharyya:2008xc}, and later in
\cite{HydroGrav,Caldarelli:2008ze}, where it was explicitly shown
that a gravitational geometry satisfies perturbatively Einstein-AdS
gravity to any order as long as the associated holographic stress
tensor $T^{\mu\nu}$ (read from the AdS/CFT dictionary) has vanishing
spacetime divergence, $\nabla_\nu T^{\mu\nu}=0$. One recognizes the
latter equations as those that govern fluid dynamics. At leading
order the stress tensor is that of a perfect fluid; in the
next-to-leading order in the perturbation, $T^{\mu\nu}$ gets a
contribution that describes viscosity and dissipation effects; at
higher order $T^{\mu\nu}$ provides information about the fluid
relaxation timescales. Thus Einstein-AdS gravity is indeed dual to
hydrodynamics, in the appropriate regime.

\begin{figure}[ht]
 \begin{center}
 \includegraphics[width=11cm]{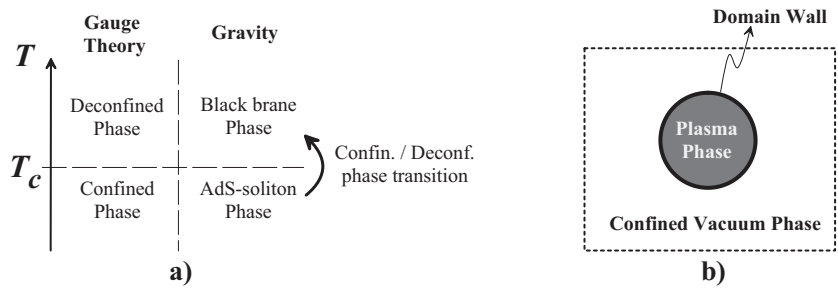}
 \caption{(a) Dominating phases in the gauge theory and in gravity. For temperatures below
 the confinement temperature $T_c$, the confined hadronic phase (AdS-soliton in gravity)
 dominates the partition function, while above $T_c$ the deconfined gluon phase (black brane in gravity)
 dominates. At $T_c$ the system suffers a first order
 confinement/deconfinement phase transition. (b) At $T_c$ and in its vicinity,
 in the dual fluid description on the boundary of the AdS gravity solution, the black hole is
 described by a plasma phase immersed in the confined vacuum phase and separated by a domain wall
 with surface tension.}\label{Fig:PlasmaPhases}
 \end{center}
\end{figure}

The hydrodynamic description of gravity
\cite{Bhattacharyya:2008jc,Bhattacharyya:2008xc,HydroGrav,Caldarelli:2008ze}
has support on the AdS/CFT duality, relating type IIB string theory
on AdS$_5\times S^5$ with $\mathcal{N}=4$ Super Yang-Mills (SYM)
gauge theory. SYM differs considerably from QCD. For example, as
opposed to SYM, QCD is non-conformal, non-supersymmetric (non-SUSY)
and has both a confined and a deconfined phase. Thus, the holy graal
of the field is to find a string/QCD duality, which would allow one
to study hard non-perturbative phenomena in QCD through a
weak-coupling perturbative analysis of the dual string system, and
vice-versa. So far this programme has not been completed (see
\cite{KlebanovMaldacena,Gubser:2009md} for discussions), however
some gravity/gauge dualities are known where the gauge theory shares
some important common features with QCD. The simplest example is the
Scherk-Schwarz (SS) compactification of a $5$-dimensional ($5d$) CFT
which yields a $4d$ non-conformal, non-SUSY gauge theory with a
confinement/deconfinement phase transition
\cite{Witten:1998zw}\footnote{SS theory has a transition from a
hadronic phase to a gluon phase. To have instead a quark-gluon phase
at high temperature we need to add fundamental matter to the model.
This can be done through the introduction of probe D6-branes
\cite{Kruczenski:2003uq} or D8-branes \cite{Sakai:2004cn}. The
latter system is known as holographic QCD or Sakai-Sugimoto model.
Currently, these gauge systems are the closest to QCD we can have
with a theory that has a gravity dual \cite{Gubser:2009md}.}. The
original CFT is $5d$ maximally SUSY SYM theory that describes field
excitations living in a stack of $D4$-branes. Identifying
periodically one of the worldvolume directions of the $D4$-branes,
imposing anti-periodic boundary conditions for the fermions along
this direction, and finally dimensionally reducing along this
compact direction one gets the desired SS gauge theory. The
gravitational dual description of this system is obtained by taking
the appropriate decoupling limit of the geometry describing the
near-extremal $D4$-branes with the compact SS worldvolume direction.
Because of the presence of this SS direction there are two
solutions: one is a black brane (deconfined gluon phase) and the
other one is the AdS-soliton (confined hadronic phase)
\cite{Horowitz:1998ha}. The solution that dominates the partition
function is the one that minimizes the free energy of the system.
One finds that there is a critical temperature $T_c$ - the
confinement temperature - above (below) which the black brane
(AdS-soliton) minimizes the free energy \cite{Witten:1998zw}. The
confinement temperature is the one where the Euclidean time circle
has the same length as the SS circle. At this confinement
temperature the two phases can co-habit in equilibrium separated by
a domain wall. Not less important, at this temperature one can have
a confinement/deconfinement phase transition in the gauge theory
which corresponds in the dual gravity side to a phase transition
between the thermal AdS-soliton and the black brane phases
\cite{Witten:1998zw} (in the context of global AdS backgrounds the
transition between thermal global AdS and the Schwarzchild black
hole solution is known as Hawking-Page phase transition). Some of
these properties are schematically represented in
Fig.~\ref{Fig:PlasmaPhases}.a). The black hole, solution of the SS
compactification of AdS$_5$ gravity on a circle, that interpolates
between the black brane phase and the AdS-soliton confined phase
with a domain wall in between is still not known. But in Ref.
\cite{Aharony:2005bm} a numerical solution was found which describes
an infinite planar domain wall separating the black brane in one
side from the confined AdS-soliton on the other, at the confinement
temperature. This solution is expected to describe approximately the
near-horizon geometry of the above mentioned black hole solution in
the limit where the black hole is large.

Strong arguments suggest that at high energy densities the SS
compactification of $5d$ CFT also has a long wavelength effective
hydrodynamical description \cite{Aharony:2005bm}. Indeed, as
described two paragraphs above this is certainly true for a CFT, and
a similar proof (although necessarily more complicated) should
follow similarly for a SS compactification of a CFT. Then, the
corresponding gravity/gauge  duality, asserts that $3d$ fluid
dynamics provides an effective theory describing the SS
compactification of $AdS_5$ gravity in the long wavelength regime
\cite{Aharony:2005bm,Lahiri:2007ae}. That is, on the boundary of a
SS compactification of AdS$_5$ (asymptoting to
$\mathcal{M}^{3}\times \calS^1$, with $\calS^1$ the distinguished SS
circle), the black hole is described by a plasma lump immersed on
the vacuum confined phase with a domain wall with surface tension
separating the two phases. This is schematically represented in
Fig.~\ref{Fig:PlasmaPhases}.b). In this description, the black hole
horizon maps to the full plasma lump bulk, not to its boundary. The
SS circle plays a minor role on the fluid description (meaning that
the plasma lumps are translationally invariant along this direction)
but as we enter through the radial holographic direction into the
bulk, the SS circle must shrink to zero size at the horizon where
the domain wall is. This implies that the topology of the
corresponding event horizon is given by the fibration of the SS
circle $\calS^1$ over the plasma lump geometry, with the circle
shrinking to a point on the boundary. So, \eg a plasma ball with a
disk topology $D^2$ corresponds in the bulk to a black hole with
horizon topology $S^3$, and a plasma ring with topology $S^1\times
I$ is dual to a black ring with topology $S^1\times S^2$
\cite{Lahiri:2007ae}. To leading order, \ie without dissipation, the
plasma lump is described by a perfect fluid stress tensor with an
equation of state characterizing the fluid from which the gauge
theory is ``made of''. The domain wall contributes with a boundary
term, proportional to its surface tension, to the stress tensor.
Finally, the system is calibrated in such a way that the confined
phase exterior to the plasma lump is vacuum with zero pressure.
These plasma balls and plasma rings were studied in great detail in
\cite{Lahiri:2007ae} with an emphasis on the AdS$_5$ case, and the
full phase diagram for balls and rings in AdS$_6$ was obtained more
recently in \cite{Bhardwaj:2008if}.
%

\subsection{A brief summary of our results}
\begin{figure}[ht]
 \begin{center}
 \includegraphics[width=6cm]{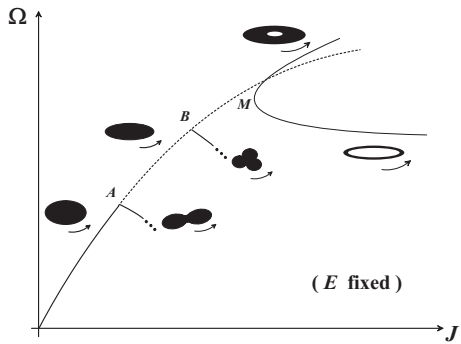} 
 \caption{Schematic phase diagram with angular velocity {\it vs} angular momentum of stationary solutions at fixed energy.
 The rotating plasma balls and plasma rings (``fat'' and ``thin'' rings merge at point $M$) phases were already studied in Ref. \cite{Lahiri:2007ae}.
 Here we find that above a critical rotation (represented by points $A$, $B$, ...), plasma balls become unstable against $m$-lobed
 perturbations (the unstable balls are represented by a dash line). The marginal point where the instability sets in is a bifurcation point
 to a new branch of non-axisymmetric $m$-lobed plasma lumps. In the simplest $m=2$ case, this is a rotating plasma peanut.
 This phase diagram summarizes some of our main results, accurately represented in Fig.~\ref{Fig:PhDiagPeanut}.}
 \label{Fig:PhaseDiagInt}
 \end{center}
\end{figure}

In the present study we analyze the stability of these plasma lumps,
with an emphasis on plasma balls. We find that rotating plasma balls
are unstable against $m$-lobed perturbations (with $m$ being the
azimuthal number of the perturbation) if their rotation is higher
than a critical value. We further find that the marginal unstable
mode is a bifurcation point to a new branch of stationary solutions
in the phase diagram of solutions. This new phase describes
$m$-lobed plasma lumps. In the simplest $m=2$ case we have a 2-lobed
configuration which presumably (if we draw from experience with
classical incompressible fluids
\cite{chandra65,brownscriven,Physics.1.38}) goes over to a
peanut-like configuration for large enough angular momentum. In this
work, we shall refer to such solutions interchangeably as rotating
{\it plasma peanut}, or 2-lobed configurations. A phase diagram
including also this new family is sketched in Fig.
\ref{Fig:PhaseDiagInt}. The associated gauge/gravity duality will
then be used to predict that black holes asymptoting to a SS
compactification of AdS$_5$ should also be unstable to $m$-lobed
perturbations. The result of the current study provides a good
example of how the hydrodynamic description of gravity can provide a
powerful predictive tool to discuss black hole physics and the
associated dual gauge theory.
Note that although we restrict our analysis to $d=3$ plasmas dual to
black objects in SS AdS$_5$, our main results and conclusions should
extend to any $d$-dimensional plasma lump dual to SS AdS$_{d+2}$
black objects, and to other gauge/gravity dualities with confined
and deconfined phases and admitting a fluid description (see however
discussion in Section \ref{sec:BallBH}).

The instability of rotating plasma balls could have been guessed from classical works and
experiments with rotating fluids \cite{plateau,brownscriven,chandra65,hilleaves,lebedev}.
A similar reasoning was recently used \cite{Caldarelli:2008ze,Caldarelli:2008mv}
to relate the Gregory-Laflamme gravitational instability of black strings and
branes to the Rayleigh-Plateau instability of plasma tubes. In the rest of this
introduction we describe with some detail the aforementioned
classical studies on rotating fluids.

The stability of {\it incompressible} $4d$ {\it non}-relativistic
fluids was studied in detail by Chandrasekhar \cite{chandra65} and
later completed by Brown and Scriven \cite{brownscriven}, for fluids
held by surface tension and in the absence of gravity. The analysis
was extended in Ref. \cite{Cardoso:2006sj} to a general number of
spacetime dimensions. In short these works found the following.
Start with a static fluid ball and slowly add rotation. The fluid
ball starts to flatten at the poles. One finds a critical rotation
frequency $\Omega_{c}$, after which two or more families can
co-exist. One of the families is the axisymmetric one. Proceeding
along this family while increasing rotation, one finds a second
critical frequency $\Omega_*$ at which the ball ``pinches'' at the
origin, and becomes a doughnut like configuration. The axisymmetric
family was found to be {\it unstable} to small perturbations after
the point $\Omega_c$. The other family that bifurcates off at
$\Omega_c$ is a ``two-lobed'' configuration (see Figs. 1 and 2 in
\cite{Cardoso:2006sj}). Close enough to the bifurcation point, this
family is {\it stable} to small perturbations. There are other
families, three-lobed, four-lobed, etc, branching off at different
points as one increases rotation. These seem to be always unstable.
Recent accurate experiments \cite{hilleaves} have confirmed the
existence, bifurcation points, and stability properties of these
families. This kind of evolution diagram was found to hold also in
gravitationally bound (Newtonian) objects with the same qualitative
behavior: the Mac-Laurin configuration being the axisymmetric family
and the Jacobi sequence, a tri-axial ellipsoid, branching off at the
bifurcation point \cite{Chandrasekhar:1992pr}.

The plan of the rest of the paper is the following.
Section~\ref{sec:Hydrodynamics} reviews the relativistic
hydrodynamic equations, the equilibrium conditions, equation of
state, and the conserved charges, that govern a Scherk-Schwarz
plasma in a $3d$ Minkowski background (we try to be self-contained).
Section~\ref{sec:ballsRings} then discusses the properties of the
axisymmetric lumps of the theory, namely the plasma balls and plasma
rings, and discusses the regime of validity of the hydrodynamic
description. In Section~\ref{sec:stability} we discuss in detail the
stability of plasma balls, the stability of plasma rings is deferred
to Appendix~\ref{sec:StabRings}. In Section~\ref{sec:bifphase} we
find that the marginal stable mode found in the previous section is
a bifurcation point to a new branch of lobed lumps.
Section~\ref{sec:BallBH} discusses the consequences of these
findings to the phase diagram of black hole solutions in SS AdS
gravity.

 \setcounter{equation}{0}
\section{Relativistic hydrodynamic equations \label{sec:Hydrodynamics}}

Here we review the relativistic hydrodynamic equations governing a
Scherk-Schwarz plasma in a $3d$ Minkowski background. We also derive
the dissipation contribution to the hydrodynamic equations which are
not easy to find in the literature. We will be interested in plasma
configurations in mechanical and thermodynamic equilibrium. We
follow closely \cite{Lahiri:2007ae,Caldarelli:2008mv}.

\subsection{Relativistic hydrodynamic equations \label{sec:HydroEqs}}

Fluid mechanics is an effective description at long distances, valid
 when the fluid variables vary on distance scales that are large
compared to the mean free path $l_\mathrm{mfp}$ of the system. As a
consequence it is natural to expand the stress tensor in powers of
derivatives of the four-velocity $u^{\mu}$. To zeroth order in the
derivative expansion, Lorentz invariance and the correct static
limit uniquely determine that the stress tensor is the sum of the
perfect fluid plus boundary contributions,
\begin{equation}
T^{\mu\nu}_{\rm perf}=\lp\rho+P\rp u^\mu u^\nu+P g^{\mu\nu}
\,,\qquad T^{\mu\nu}_{\rm bdry}= -\sigma h^{\mu\nu}|\partial
f|\,\delta(f)\,.\label{GenLumpTuv}
\end{equation}
Here, $u^{\mu}$ is the fluid velocity, $\rho$, $P$ and $\sigma$ are
the density, pressure and surface tension of the fluid. The fluid
boundary is defined by $f(x^{\mu})=0$, it has unit spacelike normal
$n_{\mu}=\partial_{\mu} f/|\partial f|$, and
$h^{\mu\nu}=g^{\mu\nu}-n^\mu n^\nu$ is the projector onto the
boundary.

The first subleading order of the derivative expansion introduces
dissipation effects in the problem. Lorentz invariance and the
physical requirement that entropy variation is non-negative demands
that the dissipation stress tensor is
\cite{Bhattacharyya:2007vs}\footnote{At leading order the entropy
current density is $(J^\mu_S)_\mathrm{perf}=s u^\mu$ and is
conserved. In the first subleading order, one gets the extra
dissipative contribution $(J^\mu_S)_\mathrm{diss} =
\frac{q^\mu}{\tloc}$. The entropy current density
$J^\mu_S=(J^\mu_S)_\mathrm{perf}+(J^\mu_S)_\mathrm{diss}$ is no
longer conserved and satisfies $\tloc\nabla_\mu J^\mu_S =
\frac{q^\mu q_\mu}{\kappa \tloc}  + \zeta \theta ^2 + 2 \eta
\sigma_{\mu \nu} \sigma^{\mu \nu}>0$ as long as $\eta, \zeta$ and
$\kappa$ are positive parameters, as we assume \cite{Lahiri:2007ae}.
In equilibrium, $\nabla_\mu J^\mu_S$ must vanish. It follows that,
$q^\mu$, $\theta$ and $\sigma^{\mu \nu}$ must vanish in
equilibrium.}
\begin{equation}\label{extraTvisc:eq}
  T^{\mu\nu}_\mathrm{diss} = -\zeta \vartheta P^{\mu\nu} -
  2\eta\sigma^{\mu\nu} + q^\mu u^\nu + u^\mu q^\nu\,,
\end{equation}
where $\zeta$ is the bulk viscosity, $\eta$ is the shear viscosity,
$\kappa$ is the thermal conductivity, and (for $d=3$)
\begin{eqnarray}\label{fluidtensors:eq}
  a^\mu &=& u^\nu \nabla_\nu u^\mu\,,\qquad  \vartheta = \nabla_\mu u^\mu, \nonumber \\
  \sigma^{\mu\nu} &=& \frac{1}{2} \lrp{P^{\mu\lambda} \nabla_\lambda u^\nu
                   + P^{\nu\lambda} \nabla_\lambda u^\mu}
                   - \frac{1}{2} \vartheta P^{\mu\nu}\,,\qquad
   q^\mu = -\kappa P^{\mu\nu} (\p_\nu\tloc + a_\nu\tloc)\,,
\end{eqnarray}
are the acceleration, expansion, shear viscosity tensor, and heat
flux, respectively. The last equation is the relativistic Fourier
law.

The hydrodynamic equations describe the conservation of the stress
tensor,
\begin{equation}
\nabla_{\mu}\lp T^{\mu\nu}_{\rm perf}+T^{\mu\nu}_{\rm
bdry}+T^{\mu\nu}_{\rm diss}\rp =0\,. \label{consTuv:diss}
\end{equation}
It then follows that in the presence of dissipation, the
relativistic continuity, Navier-Stokes and Young-Laplace equations,
are respectively given by\footnote{To go from (\ref{consTuv:diss})
into (\ref{continuity})-(\ref{YoungLap}) note that
(\ref{consTuv:diss}) reduces to a volume and a boundary
contributions. The latter gives the Young-Laplace equation while the
former yields the Continuity and Navier-Stokes equations. To get
these we use $\nabla_{\mu} \Theta(-f)= -\lp\partial_{\mu}f \rp
\,,\nabla_{\mu}\delta(f)= -|\partial f| n_{\mu}\, \delta(f)$,
$u_{\nu}u^{\nu}=-1$, $h^{\mu\nu}n_{\mu}=0$, $P^{\mu\nu}u_{\mu}=0$,
$u_{\nu}a^{\nu}=0$, $u^\mu n_{\mu}=0$, $q^\mu u_{\mu}=0$,
$\sigma^{\mu\nu}u_{\mu}=0$, $u^\mu u_{\mu}=-1$, $n^\mu n_{\mu}=1$,
and $\nabla_\alpha g^{\mu\nu}=0$. We also use properties of the
type: $n_{\nu}n^{\nu}=1$ implies
$n_{\nu}n^{\mu}\nabla_{\mu}n^{\nu}=0$.}
\begin{eqnarray}
&& \hspace{-1.5cm}u^{\mu}\nabla_{\mu}\rho + (\rho+ P) \vartheta=
\zeta\vartheta^2-q^\mu a_\mu -\nabla_\mu q^\mu
+2\eta\sigma^{\mu\nu}\nabla_\mu u_\nu\,,
\label{continuity} \\
&& \hspace{-1.5cm} (\rho+ P) a^\nu = -P^{\mu\nu} \nabla_\mu  P
+\zeta\lrp{P^{\mu\nu}\nabla_\mu\vartheta +\vartheta u^\mu\nabla_\mu
u^\nu}
 +2\eta\lrp{\nabla_\mu\sigma^{\mu\nu}-u^\nu\sigma^{\mu\alpha}\nabla_\mu u_\alpha}   \nonumber \\
&& \hspace{0.8cm} -\lrp{q^\mu\nabla_\mu u^\nu +\vartheta q^\nu
+u^\mu\nabla_\mu q^\nu -q^\mu a_\mu u^\nu}\,, \qquad {\rm with}\quad
P^{\mu\nu}\equiv g^{\mu\nu}+u^\mu u^\nu\,,
 \label{NavierS}  \\  
&& \hspace{-1.5cm}
 \left[ P - \zeta \vartheta+2\eta \lrp{\frac{1}{2}\vartheta
 +u^\mu n^\alpha\nabla_\alpha n_\mu}\right]^<_>=\sigma K\,, \qquad {\rm with}\quad  K\equiv h_{\mu}^{\:\:\nu}\nabla_{\nu}  n^{\mu}\,,
 \label{YoungLap}            
 \end{eqnarray}
where $P^{\mu\nu}$ is the projector onto the hypersurface orthogonal
to $u^{\mu}$, $K$ is the boundary's extrinsic curvature, and
$[Q]^<_>\equiv Q_<-Q_>$ is the jump on a quantity $Q$ when we cross
the boundary from the interior into the exterior of the plasma (we
will be interested in the case where the plasma object is immersed
in vacuum; then the outside contribution in the lhs of
(\ref{YoungLap}) vanishes). In the derivation of Eq.
(\ref{YoungLap}), the constraint that the fluid velocity must be
orthogonal to the boundary normal is used (this guarantees that the
fluid is confined inside the boundary),
\begin{eqnarray}
 u^\mu n_{\mu} = 0 \,. \label{constraintYL}
\end{eqnarray}
For a conformal plasma it is well-known that the bulk viscosity
coefficient vanishes, $\zeta=0$,  and that the ratio of the shear
viscosity $\eta$ to the entropy density $s$ of the plasma is
$\frac{\eta}{s}=\frac{1}{4\pi}$. However, our system is described by
a Scherk-Schwarz plasma that is not conformal and the dissipation
coefficients satisfy the relations $\zeta>0$,
$\frac{\eta}{s}>\frac{1}{4\pi}$ and $\kappa>0$.

\subsection{Hydrodynamical and thermal equilibrium conditions. Equation of state \label{sec:HydroThermoEoS}}

In this subsection we briefly review some general results derived in
Ref. \cite{Caldarelli:2008mv} valid for equilibrium and general
(non-)axisymmetric plasma configurations. We focus our discussion on
fluids in a $3d$ Minkowski background\footnote{We use polar
coordinates $(t,r,\phi)$, and the non-vanishing affine connections
are $\Gamma^r_{\phi\phi}=-r$ and
$\Gamma^\phi_{r\phi}=\Gamma^\phi_{\phi r}=1/r$.} with stationarity
timelike Killing vector $\xi=\partial_t$ and spacelike Killing
vector $\chi=\partial_\phi$, but the results of
\cite{Caldarelli:2008mv} were derived for a general background
geometry.

A fluid with local entropy density $s$ and local temperature ${\cal
T}$ satisfies the Euler relation and the Gibbs-Duhem relation given,
respectively, by
\begin{equation}
\rho+P ={\cal T} s\,, \qquad dP=sd{\cal T} \,, \label{Euler}
\end{equation}
where the latter relation follows from differentiating the former
relation and use of the first law of thermodynamics. Using these
relations and demanding hydrodynamic (\ie mechanical) equilibrium we
find that the plasma must also be in thermodynamic equilibrium
\cite{Caldarelli:2008mv}. So, equilibrium plasma configurations must
satisfy the hydrodynamic equations discussed in the previous
subsection with vanishing subleading dissipation and diffusion
contributions. It then follows that any stationary fluid
configuration with local temperature $\T$ must have a velocity given
by
\begin{equation} u=\frac\T T \lp \xi-\Omega \chi\rp\,. \label{velocityfield} \end{equation}
with constant $T$ and $\Omega$. So stationary configurations are
rigidly rotating equilibrium solutions with constant plasma
temperature $T$ related to the local temperature $\T$ by the Lorentz
factor (we use $u^2=-1$),
\begin{equation} \gamma=\frac\T T= \left[-(\xi-\Omega\chi)^2\right]^{-1/2}\,,
\label{localtemp} \end{equation}
which is the redshift factor relating measurements done in the
laboratory and comoving frames.

Combining the Euler relation (\ref{Euler}) and the Young-Laplace
equation (equation (\ref{YoungLap}) without the dissipative terms),
we can relate the plasma temperature $T$ to a combination of several
magnitudes at the fluid surface,
\begin{equation}\label{TsigmaK} T=\frac{\sigma K +\rho}{\gamma s}\,.
\end{equation}
We see that $T$ is \textit{not} simply proportional to the surface
tension or to the mean curvature, although it grows linearly with
both \cite{Caldarelli:2008mv}. For a static fluid $K$ will be
constant over the surface, but in a stationary configuration $K$ is
not a constant along the boundary. In the duality to a black hole,
$T$ corresponds to the Hawking temperature of the horizon.

A universal behavior of fluids is that they always pick boundary
configurations which reduce their potential energy for a fixed
volume. For static solutions, this implies that the area of the
fluid surface is minimized. For stationary solutions, the potential
energy not only has a surface tension term but also a centrifugal
contribution. In Ref. \cite{Caldarelli:2008mv} it was shown that this
variational principle still holds for relativistic fluids and that the
minimization problem is equivalent to maximize the plasma entropy while keeping its energy
and angular momenta fixed. In the gravitational dual system black holes satisfy the variational
principle that their entropy is extremized for fixed energy and
angular momenta. In the duality between SS-AdS black holes and fluid
lumps, the entropy, energy and angular momentum are identified on
both sides, while the temperature is mapped according to
\eqref{TsigmaK}. Note that unexpectedly (because it looks {\it a
priori} to be a contradiction), the analysis of Ref.
\cite{Caldarelli:2008mv} shows that maximization of the black hole
horizon area is equivalent, for static configurations, to
minimization of the fluid surface area. The reason for this equivalence can be traced back
to the fact that the black hole horizon is mapped
to the entire volume of the plasma and not to the plasma boundary.

The results quoted so far are independent of the equation of state
for the fluid. The gravitational/hydrodynamic duality where we frame
our analysis requires however a particular equation of state.
Indeed, we are interested in the long wavelength limit of a
Scherk-Schwarz compactification of a $4d$ CFT. The $3d$
(non-conformal) plasma that results from the dimensional reduction
of the $4d$ conformal plasma has equation of state defined by
\cite{Lahiri:2007ae}
\begin{equation}
 P=\frac{\rho-4\rho_0}{3}\,, \qquad \rho+P=\frac{4}{3}\lp \rho-\rho_0\rp
 \,,\qquad
 s=4\alpha^{1/4}
\left(\frac{\rho-\rho_0}{3}\right)^{\frac{3}{4}}\,,\qquad {\cal
T}=\left(\frac{\rho-\rho_0}{3\,
\alpha}\right)^{\frac{1}{4}}\,.\label{ConfEqState}
 \end{equation}
with $\rho_0$ and $\alpha$  constants. This equation of state is
valid in or out of equilibrium and is normalized such that the
vacuum pressure vanishes. For the SS plasma in equilibrium, it
follows from (\ref{localtemp}) and (\ref{ConfEqState}) that the
pressure and energy density satisfy the relations
\begin{equation}
 P=\frac{\rho_{*}}{3}\,\gamma^4-\rho_0\,,\qquad \rho=\rho_{*}\,\gamma^4+\rho_0\,,
\label{DpEquil}
\end{equation}
where $\rho_{*}$ is a constant.

\subsection{Conserved charges}
The constituent fluid of the plasma object has local energy density
$\rho$, pressure $P$, velocity $u^{\mu}$, local entropy density $s$,
and local temperature ${\cal T}$. These {\it local} quantities
provide the information we need to compute the thermodynamic
quantities (energy, angular momentum, entropy, temperature) of the
plasma balls and plasma rings.

To define these quantities recall that our fluid lives in a $3d$
Minkowski background with stationarity timelike Killing vector
$\xi=\partial_t$ and spacelike Killing vector $\chi=\partial_\phi$.
We can then foliate the spacetime into constant $t$ hypersufaces
$\Sigma_t$ and $\xi^\mu$ is their unit normal vector. Then, given
any Killing vector $\psi^\mu$, one can define the associated
conserved charges ${\mathcal Q}[\psi]=\int_{\Sigma_t}\!\!dV
\;T_{\mu\nu}\xi^{\mu}\psi^\nu$, where $dV$ is the induced volume
measure on $\Sigma_t$. The fluid velocity is given by
(\ref{velocityfield}), \ie $ u^{\mu}=\gamma \lp \delta^{\mu t} +
\Omega\delta^{\mu\phi} \rp$ with  $\gamma=\lp 1-r^2\,\Omega^2
\rp^{-1/2}$. The energy and angular momentum of the plasma
associated, respectively, with the Killing vectors $\xi$ and $\chi$
are then
\begin{eqnarray}
&& E=\int_V dV\left[ \gamma^2\lp \rho+
r^2\,\Omega^2P \rp   
 +\sigma |\partial f|\,\delta(f)
\right]  \,,\nonumber\\
&& J=\int_V\, dV\, r^2\,\Omega \gamma^2\lp\rho+P\rp
  -\sigma\int_{\Sigma_t} dV \, h_{\mu\nu}\xi^\mu\chi^\nu|\partial
f|\delta(f) \,.
 \label{Var:charges}
\end{eqnarray}
Note that for axisymmetric solutions (plasma balls and plasma rings)
the boundary term in $J$ proportional to $\sigma$ vanishes. It is
however present for non-axisymmetric solutions where $\chi\cdot
n\neq 0$ (\ie when the fluid boundary is not invariant under the
action of $\chi$) \cite{Caldarelli:2008mv}.
The total entropy of the fluid is the conserved charge associated to
the entropy density current $su^\mu$,
\begin{eqnarray}
S=-\int_V dV\,(\chi\cdot u) s =\int_V\, dV \gamma s
\,.\label{Var:Entropy}
\end{eqnarray}

 \setcounter{equation}{0}
\section{Equilibrium solutions: rigidly rotating plasma balls and rings \label{sec:ballsRings}}

There are three families of axisymmetric rigidly rotating
equilibrium configurations in a $3d$ Minkowski background: plasma
balls, plasma rings and plasma tubes. The later were already
analyzed in a previous paper and do not interest us here
\cite{Caldarelli:2008mv}. The plasma balls and rings were discussed
in detail in \cite{Lahiri:2007ae}. Because we will later study the
stability of these solutions we review were their properties,
following closely \cite{Lahiri:2007ae}.

Consider plasma configurations in a $d=3$  Minkowski background
parametrized by coordinates $(t,r,\phi)$. The axisymmetry
requirement demands that the boundaries of the plasma depend only on
$r$. Each boundary is thus defined by the condition ($j$ specifies a
particular boundary in the case where more than one is present)
\begin{equation} f(r)= r-R_j=0\,,
\end{equation}
and has unit normal $n_\mu=\frac{\partial_\mu f}{|\partial
f|}=\delta_{\mu r}$.  Its extrinsic curvature
$K=h_{\mu}^{\:\:\nu}\nabla_{\nu} n^{\mu}$ is $K=\frac{1}{R_j}$.

Following \cite{Lahiri:2007ae}, it is convenient to frame our
discussion in terms of the dimensionless variables,
\begin{equation}\label{newvars}
    \tO = \frac{\sigma\Omega}{\rho_0} \,,   \qquad
    \z = \frac{\rho_0 r}{\sigma} \,,   \qquad
    v = \Omega r = \tO \z \,,
\end{equation}
and also to use dimensionless thermodynamic quantities,
\begin{equation}\label{redThermo}
  \tE = \frac{\rho_0 E}{\pi\sigma^2} \,,\quad
  \tJ = \frac{\rho_0^2 J}{\pi\sigma^3} \,,\quad
  \tS = \frac{\rho_0^{5/4}S}{\pi\alpha^{1/4}\sigma^2} \,,\quad
  \tT = T\lrp{\frac{\alpha}{\rho_0}}^{1/4}\,.
\end{equation}

We now consider the properties of plasma balls and rings.

\subsection{Plasma Balls}
Plasma balls are characterized by having a single axisymmetric outer
surface at $r=\ro$ and by $P_>=0$. Using  equation of state
(\ref{ConfEqState}), the Young-Laplace boundary condition
(\ref{YoungLap}) reads
\begin{equation}\label{Ball:bc}
  \rho(\ro) = 4\rho_0 + \frac{3\sigma}{\ro}\,.
\end{equation}
Plasma balls in equilibrium must satisfy the equation of state (\ref{DpEquil}) and obey the boundary condition (\ref{Ball:bc}).
This implies that
\begin{equation}\label{Ball:sol}
  \frac{\rho(v)-\rho_0}{3\rho_0} \lrp{1-v^2}^{2}
       = \lrp{1+\frac{\tO}{\vo}}\lrp{1-\vo^2}^{2}
       \equiv g_+(\vo)\,.
\end{equation}
The range of $v$ is $[0,1]$ and $\rho(v)-\rho_0$ is always positive
for the plasma ball. This is in agreement with the requirement that
the local temperature defined in (\ref{ConfEqState}) must be
positive. After using the equation of state (\ref{ConfEqState}) and
(\ref{Ball:sol}) we get for the local plasma temperature,
\begin{equation}\label{Ball:locTemp}
  \tloc = \gamma\lrp{\frac{\rho_0 g_+(\vo)}{\alpha}}^{1/4}\,.
\end{equation}
For later use note that the constant $\rho_{*}$ introduced in
(\ref{DpEquil}) is related to $\rho_0$ by
\begin{equation}\label{TildeRHO:rho0}
\rho_{*}=3\rho_0 g_+(\vo) \,.
\end{equation}
This relation follows from replacing (\ref{DpEquil}) in the
Young-Laplace equation.

Use of (\ref{Ball:sol}) and (\ref{ConfEqState}) in
(\ref{Var:charges})-(\ref{redThermo}) yields for the dimensionless
energy, angular momentum and entropy of the plasma ball,
\begin{equation}\label{Ball:charges}
  \tE = \frac{4\vo^2-\vo^4+5\tO \vo-\tO \vo^3}{\tO^2}
\,, \qquad
  \tJ = \frac{2\vo^4+2\tO \vo^3}{\tO^3}
\,,\qquad
  \tS = \frac{4\vo^2}{\tO^2}\sqrt{1-\vo^2}
         \lrp{1+\frac{\tO}{\vo}}^{3/4},
\end{equation}
while the dimensionless temperature and dimensionless angular
velocity of the plasma balls are
\begin{equation}\label{Ball:tempAngV}
  \tT = \pdiffc[\tJ]{\tE}{\tS}=  [g_+(\vo)]^{1/4}, \qquad
  \tO = \pdiffc[\tS]{\tE}{\tJ}.
\end{equation}
Note that the plasma ball temperature is the redshifted local
temperature, $T={\cal T}/\gamma$, in agreement with the discussion
associated with (\ref{localtemp}). The plasma angular velocity is
naturally the same as the fluid one with no associated Lorentz
factor.

\subsection{Plasma Rings}
These have an axisymmetric inner surface at $r=\ri$ (where $P_<=0$),
in addition to the outer surface at $r=\ro$ (where $P_>=0$). Using
equation of state (\ref{ConfEqState}), the Young-Laplace equation
yields the pair of boundary conditions,
\begin{equation}\label{Ring:bc}
  \rho(\ro) = 4\rho_0 + \frac{3\sigma}{\ro}\,,\qquad \rho(\ri) = 4\rho_0 - \frac{3\sigma}{\ri}\,.
\end{equation}
Plasma rings in equilibrium must also satisfy the equation of state
(\ref{DpEquil}) and obey the boundary conditions (\ref{Ring:bc}).
This means that rings must satisfy the pair of equations
\begin{eqnarray}\label{Ring:sol}
 \frac{\rho(v)-\rho_0}{3\rho_0}\lrp{1-v^2}^{2}
       &=& \lrp{1+\frac{\tO}{\vo}}\lrp{1-\vo^2}^{2}
       \equiv g_+(\vo)\nonumber\\
       &=& \lrp{1-\frac{\tO}{\vi}}\lrp{1-\vi^2}^{2}
       \equiv g_-(\vi)\,.
\end{eqnarray}
Note that $\rho(v)-\rho_0$, and thus the local temperature are
non-negative as long as $\vi \geq \tO$. This system can be satisfied
only when
\begin{equation}\label{Ring:gConstraint}
 g_+(\vo)=g_-(\vi) \,.
\end{equation}
This condition  constrains the three variables $\vo$, $\vi$ and
$\tO$ as, \eg $\vi=\vi(\vo,\tO)$. An inspection of
$g_+(\vo)=g_-(\vi)$ concludes that there is a minimum $\vo$, call it
$\vo^*$, above which (\ref{Ring:gConstraint}) is valid
\cite{Lahiri:2007ae}. So, plasma rings exist only for $\vo \geq
\vo^*$. In fact there are two families of black rings. One is called
the fat plasma ring and exists for $\tO \leq\vi \leq\vi^*$ (where
$\vi^*<\vo^*$ is such that the derivative of $g_-(\vi)$ vanishes),
while the second, dubbed as thin plasma ring, exists for $\vi^*
\leq\vi \leq 1$. At $\vi=\vi^*$ the two families meet at a regular
solution.

Use of (\ref{ConfEqState}) and (\ref{Ring:gConstraint}) yields for
the local plasma temperature
\begin{equation}\label{Ring:locTemp}
  \tloc = \gamma\lrp{\frac{\rho_0 g_+(\vo)}{\alpha}}^{1/4}
        = \gamma\lrp{\frac{\rho_0 g_-(\vi)}{\alpha}}^{1/4}\,.
\end{equation}
Finally note that $\rho_{*}$ defined in (\ref{DpEquil}) is
related to $\rho_0$ by $\rho_{*}=3\rho_0 g_+(\vo)=3\rho_0
g_-(\vi)$.

Using (\ref{Ring:sol}) and (\ref{ConfEqState}) in
(\ref{Var:charges})-(\ref{redThermo}) yields for the dimensionless
energy, angular momentum and entropy
%
\begin{eqnarray}\label{Ring:charges}
    \tE &=& \frac{4(\vo^2-\vi^2) - (\vo^4-\vi^4)
                 + 5\tO (\vo+\vi) - \tO (\vo^3+\vi^3)}
             {\tO^2}\,, \qquad
  \tJ = \frac{2(\vo^4-\vi^4)+2\tO (\vo^3+\vi^3)}{\tO^3} \,,
  \nonumber\\
  \tS &=& \frac{4}{\tO^2}
        \lrpp{\vo^2\sqrt{1-\vo^2}\lrp{1+\frac{\tO}{\vo}}^{3/4}
            \!\!-\vi^2\sqrt{1-\vi^2}\lrp{1-\frac{\tO}{\vi}}^{3/4}},
\end{eqnarray}
while the dimensionless temperature and dimensionless angular
velocity of the plasma rings are
\begin{equation}\label{Ring:tempAngV}
  \tT = \pdiffc[\tJ]{\tE}{\tS}=  [g_+(\vo)]^{1/4} = [g_-(\vi)]^{1/4}, \qquad
  \tO = \pdiffc[\tS]{\tE}{\tJ}\,.
\end{equation}
Note that to determine the plasma ring family of solutions, the
constraint (\ref{Ring:gConstraint}) must be imposed. The different
equilibrium configurations are best understood by looking at a phase
diagram of solutions shown in later sections.

\subsection{The hydrodynamic regime \label{sec:HydroRegime}}
Relativistic hydrodynamics provides a good effective description of
the deconfined plasma phase of ${\cal N}=4$ Yang Mills theory
compactified down to $d=3$ on a Scherk-Schwarz circle only if
certain conditions are satisfied \cite{Lahiri:2007ae}. First,
hydrodynamics is by definition valid when the thermodynamic
quantities of the fluid vary slowly over the mean free path
$\ell_{\rm mfp}$ of the fluid. In our case $\ell_{\rm mfp}\sim  T_c
\sim \frac{\rho_0}{\sigma}$. A good estimate for the validity regime
is obtained when the maximum fractional rate of change of the fluid
local temperature, $\frac{\delta \tloc}{\tloc}{\bigl |}_{\rm
max}\sim \partial_r \ln\gamma {\bigl |}_{\rm max}$ (recall that
$\tloc=T\gamma$) is much smaller than $\ell_{\rm mfp}$. This occurs
for $\frac{\tO \vo}{1-\vo^2} \ll 1$. This condition is satisfied by
a wide range of plasma balls and rings as long as we are away from
extremality. Second, the analysis done so far and onwards assumes
that surface tension is constant, $\sigma=\sigma( T_c )$, when in
fact it is a function of the fluid temperature at the surface. This
assumption is valid when $\tloc/ T_c \sim 1$ at the boundary
surfaces. This is the case for a long range of energies and angular
momentum as long as we do not approach too much the extremal
configurations. Finally, the boundary of the plasma is treated as a
delta-like surface when in fact it has a thickness of order $ T_c
^{-1}$. So the analysis is valid when the boundary radius is
everywhere large when compared with $ T_c ^{-1}$,
$\{\ro,\ri,\ro-\ri\} \gg \frac{\sigma}{\rho_0}$. This is the case if
the plasma energy is large and again if we are away from the
extremal configurations \cite{Lahiri:2007ae}.

\setcounter{equation}{0}
\section{Stability analysis of plasma balls and rings. \label{sec:stability}}
In this section we want to consider a rigidly rotating plasma ball
in $d=3$ and address its stability when perturbed. The dynamics
of the perturbations is dictated by the hydrodynamic equations,
subject to appropriate boundary conditions. In this section we
restore light velocity factors to ease comparison between relativistic and classical results.

\subsection{Perturbations of the equilibrium solutions: inviscid relativistic case}
Perturbations take the plasma away from thermal equilibrium and
therefore viscosity and diffusion effects start to contribute. The
energy-momentum tensor of the fluid includes not only the perfect
fluid and the boundary surface tension terms (\ref{GenLumpTuv}), but
also a dissipative contribution \eqref{extraTvisc:eq}. For now, we
neglect the dissipation contribution to the fluid stress tensor.

Consider then a generic equilibrium solution described by velocity
$u^{\mu}_{(0)}=\gamma \lp \delta^{\mu t} + \Omega\delta^{\mu\phi}
\rp$ (with $\gamma=\lp 1-r^2\,\Omega^2 /c^2 \rp^{-1/2}$), pressure
and density functions as given by (\ref{DpEquil}), which we label
with the subscript $(0)$, standing for unperturbed quantities. Now,
suppose the system acted upon by a perturbation with the generic
form
\beq
 P &=& P_{(0)}+\delta P\,, \qquad  \quad
 \delta P(t,r,\phi)=\epsilon \gamma^4 \calP(r) e^{(\omega-im\Omega) t+im\phi}\,, \\
 \rho &=& \rho_{(0)}+\delta \rho \,, \qquad
 \rho_{(0)}=\frac{3}{c^2}\,P_{(0)}+4\rho_0\,,\qquad \delta
\rho(t,r,\phi)=\frac{3}{c^2}\,\delta P(t,r,\phi) \,,\nonumber\\
 u^{\mu}&=& u_{(0)}^{\mu}+\delta u^\mu \,, \qquad  \quad
\delta u^\mu(t,r,\phi) = \epsilon
 \lp U_t(r)\delta^{\mu t}+\gamma^{-3}\,U_r(r)\delta^{\mu r}+\frac{U_\phi(r)}{r}\delta^{\mu \phi} \rp e^{(\omega-im\Omega)
t+im\phi}\,, \nonumber \label{perturbation}
 \eeq
where we used the equation of state (\ref{ConfEqState}) valid also
out of equilibrium and we denote the perturbation of a quantity $Q$
as $\delta Q$. Positive real part of $\omega$ signals an instability.
After eliminating the $0^{\rm th}$ order
terms using the unperturbed hydrodynamic equations, the continuity
and the Navier-Stokes equations yield, up to first order in the
perturbation,
\begin{eqnarray}
&&c^2 u^\mu_{(0)}\nabla_{\mu}\delta\rho+c^2\delta
u^{\mu}\nabla_{\mu}\rho_{(0)} + (\rho_{(0)}c^2+P_{(0)}) \nabla_{\mu}
\delta u^\mu + (c^2\delta\rho+\delta P) \nabla_{\mu} u^\mu_{(0)} =
0\,,\nonumber\\
&&(\rho_{(0)}c^2+P_{(0)}) \lp \delta u^\mu \nabla_{\mu} u^\nu_{(0)}
+ u^\mu_{(0)} \nabla_{\mu} \delta u^\nu \rp + (c^2\delta\rho+\delta
P) u^\mu_{(0)}
\nabla_{\mu} u^\nu_{(0)} \nonumber\\
&&\qquad\quad+ \lp g^{\mu\nu}+u^\mu_{(0)} u^\nu_{(0)} \rp
\nabla_{\mu} \delta P+ \lp  \delta u^\nu u^\mu_{(0)} + u^\nu_{(0)}
\delta u^\mu\rp \nabla_{\mu} P_{(0)}=0 \,. \label{3d:NavierS}
\end{eqnarray}
which reads
\beq 0&=& \frac{3\gamma\,r\omega }{\frac{4\rho_{*}}{3}\,c^2}  \calP
+\gamma^{-3}\frac{d}{dr}(r\,U_r)+r(\omega-im\Omega)U_t+imU_\phi
\,, \label{contd3}\\
0&=& \frac{i\gamma\,\Omega}{\frac{4\rho_{*}}{3}\,c^2}
(i\omega \Omega r^2-mc^2\gamma^{-2})\calP -2\Omega^2 r \gamma^{-1}U_r-c^2\omega \,U_t\,,\label{Nad31}\\
0&=&\frac{\gamma^2}{\frac{4\rho_{*}}{3}}\calP'+\omega  U_r-2 \gamma^{3}\Omega U_\phi\,,\label{Nad32}\\
0&=& \frac{i\gamma}{\frac{4\rho_{*}}{3}\,c^2}\lp i\omega \Omega r^2-mc^2\gamma^{-2}\rp \calP
- 2\Omega r \gamma^{-1} U_r-\omega \,r\, U_\phi \,,\label{Nad33}
\eeq
where $\calP'\equiv \frac{d\calP}{dr}$. Multiplying
(\ref{Nad33}) by $\Omega$ and subtracting (\ref{Nad31}) it follows
that $U_t=\frac{r\Omega}{c^2} U_\phi$, which satisfies the
requirement that $u_{\mu}u^{\mu}=-c^2$ up to order $\epsilon$.
These equations can be used to get a second order ODE for $\calP$,
and another equation defining $U_r$ in terms of $\calP$ and its
derivative. The later is
\be U_r(r)=\frac{-2i\Omega\lpp c^2m-r^2\Omega(i\omega+m\Omega)\rpp\gamma^{2}{\cal P}(r)-c^2\omega r
{\cal P}'(r)} {\frac{4\rho_{*}}{3}r
\lpp-\omega^2r^2\Omega^2+c^2(\omega^2+4\Omega^2)\rpp}\,,\label{vasP}
\ee
while the second order ODE for $\calP$ is
\beq 0&=&- \,r^2\,c^6\gamma^{-4}\left (c^2\,\omega^2\gamma^{-2}+4c^2\Omega^2\right )\calP''-
\,r c^6 \gamma^{-4}\left (\omega^2(c^2+\Omega^2\,r^2)+4c^2\Omega^2\right )\calP'\,\nonumber\\
&+&{\biggl[}
c^8m^2(\omega^2+4\Omega^2)-\omega^2\,r^8\Omega^6(\omega-im\Omega)^2+c^4r^4\Omega^2\left (6i\omega^3\,m\,\Omega-7\omega^4+2\omega^2\Omega^2(3m^2-14)+24i\omega\,m\,\Omega^4\right )
\,\nonumber\\
&-&c^2r^6\Omega^4\left
(6i\omega^3m\Omega-5\omega^4+4\omega^2(m^2-2)\Omega^2+12i\omega\,m\,\Omega^3+4m^2\Omega^4\right
)\nonumber\\
&+&c^6r^2\left
(3\omega^4-2i\omega^3\,m\,\Omega-4\omega^2(m^2-5)\Omega^2-12i\omega\,m\Omega^3+4(8-3m^2)\Omega^4\right
){\biggr]} \calP \label{ODE}\eeq
The perturbed continuity and Navier-Stokes equations must be
supplemented by appropriate boundary conditions. For that, let us
write a general boundary disturbance as
\be r=R(t,\phi)\,,\qquad {\rm with} \quad R(t,\phi)= R_j\lp
1+\epsilon \,\chi\,e^{(\omega-im\Omega) t+im\phi}\rp\,,\qquad
\epsilon \ll 1 \,,\label{displacementd3} \ee
where $R_j$ is the unperturbed radius of the boundary. For the
plasma ball one has $R_j\equiv R_o$, while for the plasma ring we
have both the inner and outer boundaries: $R_j\equiv R_{\rm i}$ and
$R_j\equiv R_o$. For plasma rings, this reduces the possible set of
perturbations to the subsector with similar temporal and angular
deformations in both boundaries.

The first boundary condition is a kinematic condition requiring that
the normal component of the fluid velocity on the boundary satisfies
the perturbed version of (\ref{constraintYL}),
$u^{\mu}_{(0)}\,\delta n_\mu +\delta u^\mu n_{\mu}^{(0)}=0$, where
$\delta n_\mu\equiv n_\mu{\bigl |}_{R(t,\phi)}-n_{\mu}^{(0)}$ and
the unperturbed normal is $n_\mu^{(0)}\equiv n_\mu{\bigl
|}_{R_{o,\rm i}}=\delta_\mu^r$. This ensures that the fluid is
confined inside  the boundary and is also a consistency relation
between the boundary  and the velocity perturbation. To leading
order this boundary condition reads,
\be {\rm BC}\:\: {\rm I:}\qquad U_r{\bigl |}_{R_{o,\rm i}} \simeq
\omega \gamma_{o,\rm i}^4  R_{o,\rm i}  \chi_{o,\rm i} \,.
\label{3d:bc2}
\ee
The second boundary condition is a balance on the normal stress
at the boundary. This means that the pressure perturbation
must also satisfy the perturbed version of the Young-Laplace
equation (\ref{YoungLap}), namely: $\delta P_<-\delta P_>=\sigma
\delta K$. Since we have vacuum in the exterior of the plasma
configuration this reads
\beq {\rm BC}\:\: {\rm II:}\qquad \lp P_{\lessgtr}^{(0)} {\bigl
|}_{R(t,\phi)} +\delta P{\bigl |}_{R_{o,{\rm i}}} \rp-
P_{\lessgtr}^{(0)} {\bigl |}_{R_{o,{\rm i}}}= \pm \sigma \lp K{\bigl
|}_{R(t,\phi)} -K{\bigl |}_{R_{o,{\rm i}}} \rp \,,
 \nonumber\label{3d:bc1} \eeq
where the choices $\{<,+,o\}$ apply to the outer boundary and
$\{>,-,{\rm i}\}$ to the inner boundary, if present. The subscript
$r=R(t,\phi)$ means that we evaluate the expression at the perturbed
boundary $r=R(t,\phi)$ defined in (\ref{displacementd3}) and the
subscript $r=R_{o,{\rm i}}$ means evaluation at the unperturbed
boundary $r=R_{o,{\rm i}}$. $P_{\lessgtr}^{(0)}$ is computed using
(\ref{DpEquil}). The extrinsic curvature
$K=h_{\mu}^{\:\:\nu}\nabla_{\nu} n^{\mu}$ is obtained using the unit
normal of (\ref{displacementd3}),
 \be n_{\mu}=|\delta f|^{-1}\lp
-R_t'\delta_{\mu}^{\:t}+
 \delta_{\mu}^{\:r}-R_{\phi}'\delta_{\mu}^{\:\phi} \rp \,,\qquad
 |\delta f|=\lp 1-\frac{1}{c^2}R^{\prime\,2}_t+\frac{1}{r^2}R^{\prime\,2}_\phi\rp^{\frac{1}{2}}.
 \label{3d:normal:pert}
 \ee
The Young-Laplace equation
(\ref{3d:bc2}) then yields to leading order in $\epsilon$
\be {\cal P}{\bigl |}_{R_{o,\rm i}}\simeq \frac{\sigma} {R_{o,\rm
i}} \, \chi_{o,\rm i}\gamma_{o,\rm i}^{-4}
 \lpp \pm \lp \frac{1}{c^2}(\omega-im\Omega)^2 R_{o,\rm i}^2 + m^2-1\rp
 -\Sigma\,\gamma_{o,\rm i}^6 \rpp \,, 
 \label{3d:bc1End}
 \ee
where $\pm$ applies, respectively, to the outer and inner boundary
and $\gamma_{o,\rm i}\equiv \gamma{\bigl |}_{R_{o,\rm i}}$. Here, we
have defined the rotational Bond parameter $\Sigma$ which
plays an important role in this problem. It measures the competition
between centrifugal and surface tension effects and is defined by,
\be  \Sigma \equiv \frac{4\rho_{*}}{3} R_{o,\rm
i}^3\,\frac{\Omega^2}{\sigma} \label{Bond}\,.\ee
%

\subsection{Plasma balls: instability and critical rotation in the inviscid case}
We now particularize the above framework for plasma balls.
We can combine boundary conditions (\ref{3d:bc1End}) and
(\ref{3d:bc2}) in a single condition,
\be U_r(R_o)=\frac{\omega R_o^2\,\gamma_o^{8}}{\sigma\,\lp
m^2-1
 -\Sigma \gamma_o^6 +\frac{R_o^2}{c^2}(\omega-im\Omega)^2\rp }
\,\calP(R_o)\,.\label{secondBC}\ee
To summarize, (\ref{vasP}) and (\ref{secondBC}) give us a condition on $\calP$.
Together with regularity conditions at the origin, Eq. (\ref{ODE}) is then an eigenvalue problem for $\omega$.

Although we will present our numerical results in full generality,
it is insightful to compare them with the small rotation regime
where an analytical treatment is possible. In this small velocity
regime, $\Omega R_o\ll c$, (\ref{ODE}), (\ref{vasP}) and
(\ref{secondBC}) reduce respectively to
\beq &&\frac{1}{r}\frac{d}{dr}\lp r{\cal P}'\rp-\frac{m^2}{r^2}{\cal
P}=0\,, \label{NR:1}\\
&&U_r(r)=-\frac{2im\Omega{\cal P}(r)+\omega r {\cal P}'(r)}{\frac{4 \rho_{*}}{3}r \lp\omega^2 +4\Omega^2\rp }\,,  \label{NR:2}\\
&&
U_r(R_o)=\frac{\omega R_o^2}{\sigma\,\lp
m^2-1-\Sigma \rp}
\,\calP(R_o)\,.  \label{NR:3}\eeq
If we define
\be
\Sigma_\omega \equiv \frac{4\rho_{*}}{3}R_o^3\,\omega^2\,, \label{sigmaw}
\ee
we can express the analytical solution to the system (\ref{NR:1})-(\ref{NR:3}) as
\be \sqrt{\Sigma_\omega}=i\sqrt{\Sigma}\pm\sqrt{(m-1)\lp
\Sigma-m(m+1)\rp}\label{classical}\,. \ee
Thus, for $\Sigma>m(m+1)$ and $m\geq 2$, the system is unstable.
This is one of our main results: in the simplest $m=2$ case, plasma
balls become unstable against 2-lobed or peanut-like deformations
when the rotation reaches a critical value $\Omega_{\rm crit}$. For
higher rotation they become also unstable against $m$-lobed
deformations, with $m>2$. Notice that in the non-relativistic regime
the density is approximately a constant and equal to $4\rho_*/3$
(see footnote~\ref{foot}). Thus, our result (\ref{classical}) is
precisely the well-known result for inviscid, incompressible fluids
of density $\rho=4\rho_*/3$ \cite{ViscClassic}.

\begin{figure}[h!tbp]
\begin{center}
\begin{tabular}{ll}
\epsfig{file=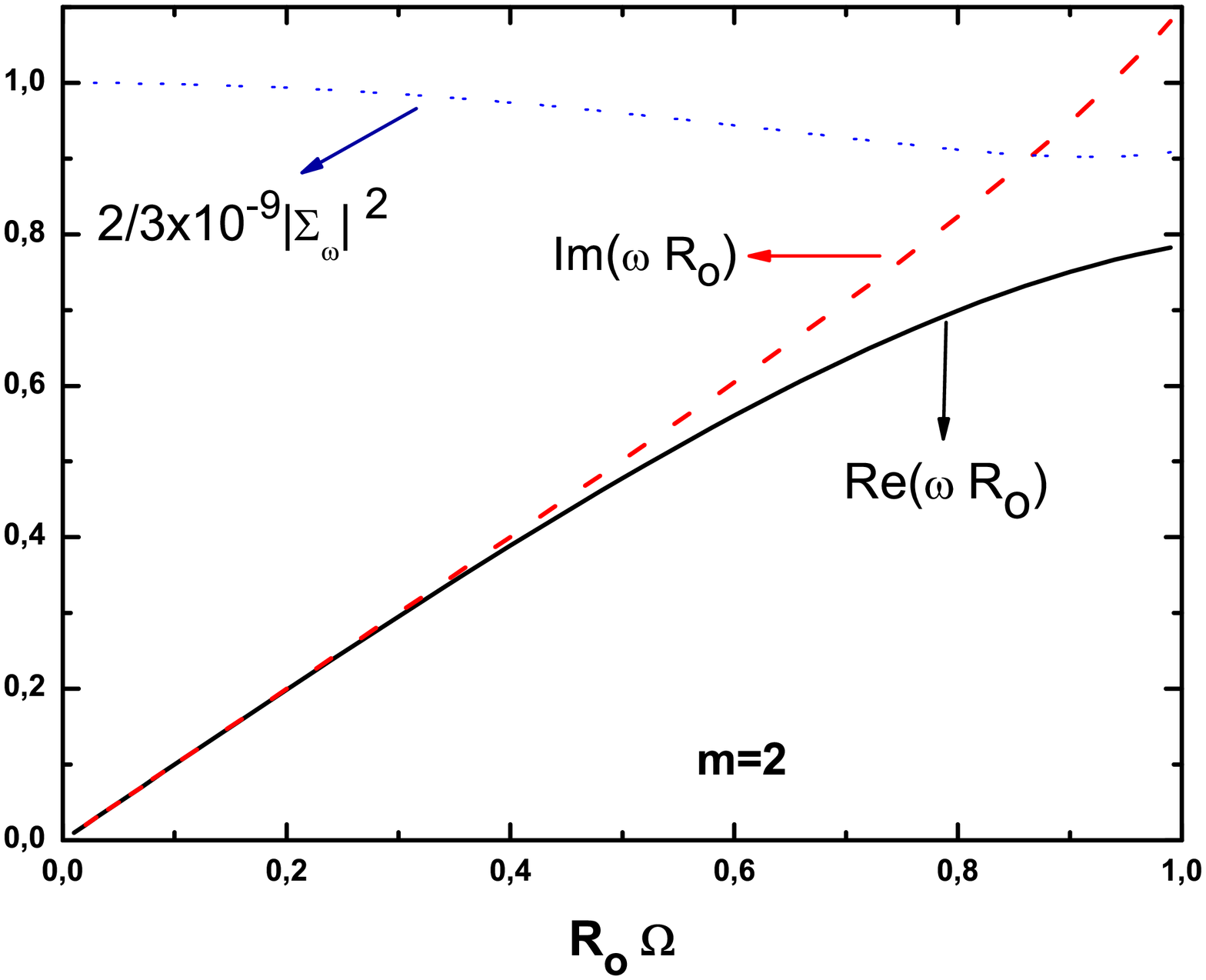,width=8cm,angle=0}&\epsfig{file=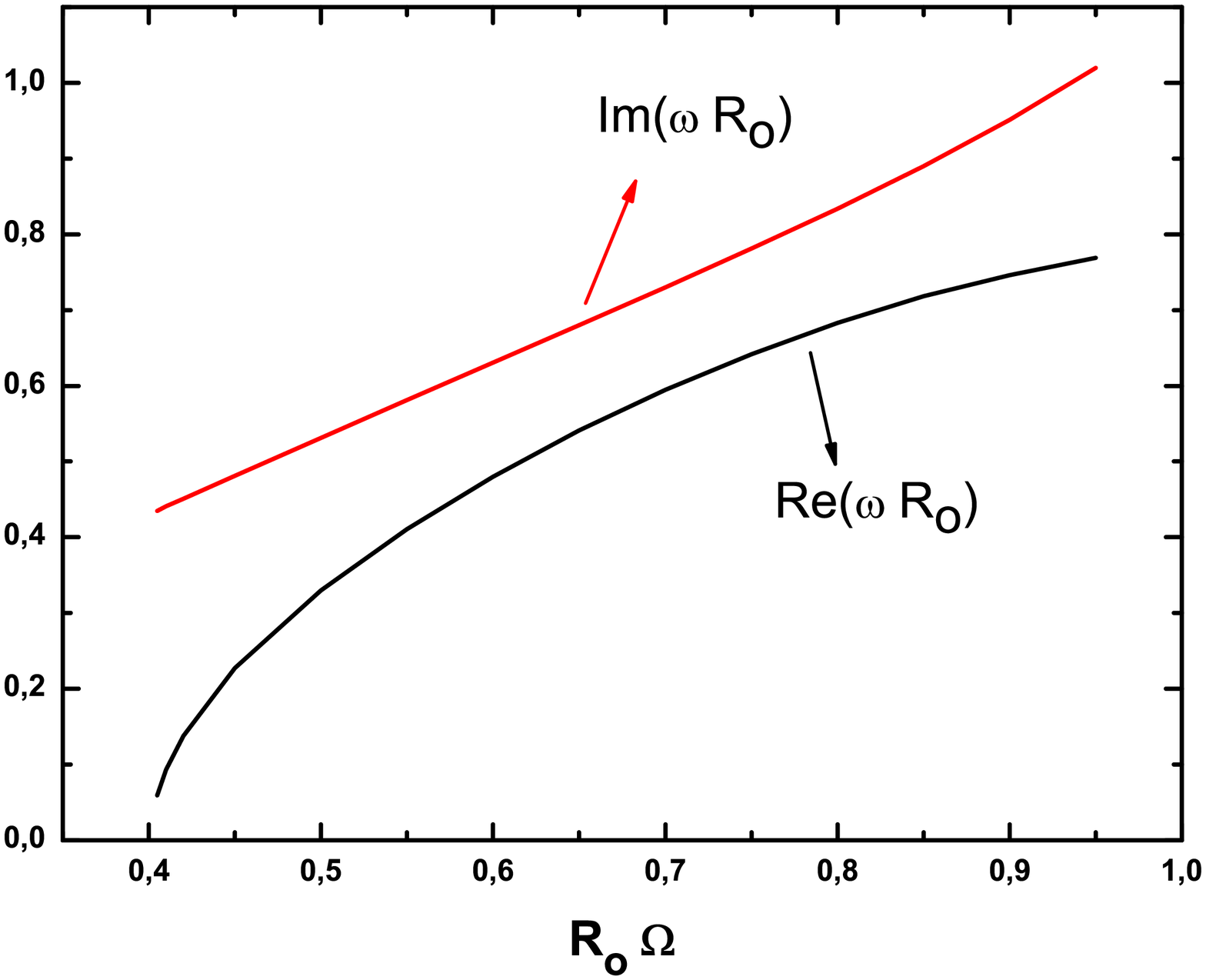,width=8cm,angle=0}
\end{tabular}
\end{center}
\caption{Left Panel: Details of instability for the $m=2$ mode. We
are fixing $\Sigma=\frac{4}{3}\times 10^9$, $\sigma=10^{-7}$. Classically, \ie for small
rotation rates, the quantity $\Sigma_\omega$ is a constant, and both
the real and imaginary part of the characteristic frequency $\omega$
lie on the straight line, \ie directly proportional to $\Omega$. The
stability region lies in a range of very small $\Omega R_o$, not
visible in the figure. Right Panel: Details of the instability for
the $m=2$ mode, this time with parameters chosen such that the
instability sets in at large $\Omega R_o$. In this case, the
threshold is around $\Omega R_o/c \sim 0.402$, the parameters are
$\frac{4\rho_{*}R_o}{3\sigma}=24$. Classically, the threshold
would be at at $\Omega R_o/c=0.5$.} \label{fig:instm2}
\end{figure}

Our numerical results for the relativistic system (\ref{ODE}),
(\ref{vasP}) and (\ref{secondBC}), are depicted in Fig.
\ref{fig:instm2} for the $m=2$ case. For small rotation rates, they
are in perfect agreement with the non-relativistic limit
(\ref{classical}).
We find a (in any case small) deviation from the classical
prediction only when $\Omega R_o/c$ approaches unity. In the right
panel, we show a case where the threshold rotation frequency at
which an instability sets in is rather large. For these values we
get a threshold of approximately $\Omega R_o/c \sim 0.402$, still in
good agreement with the classical result $\Omega R_o/c =0.5$, as
given by (\ref{classical}) for these values
($\frac{4\rho_{*}R_o}{3\sigma}=24$). We were not able to find
unstable modes for $m=1$, in agreement with the classical result
(\ref{classical}) for small rotations.

\begin{figure}[h!tbp]
\begin{center}
\begin{tabular}{l}
\epsfig{file=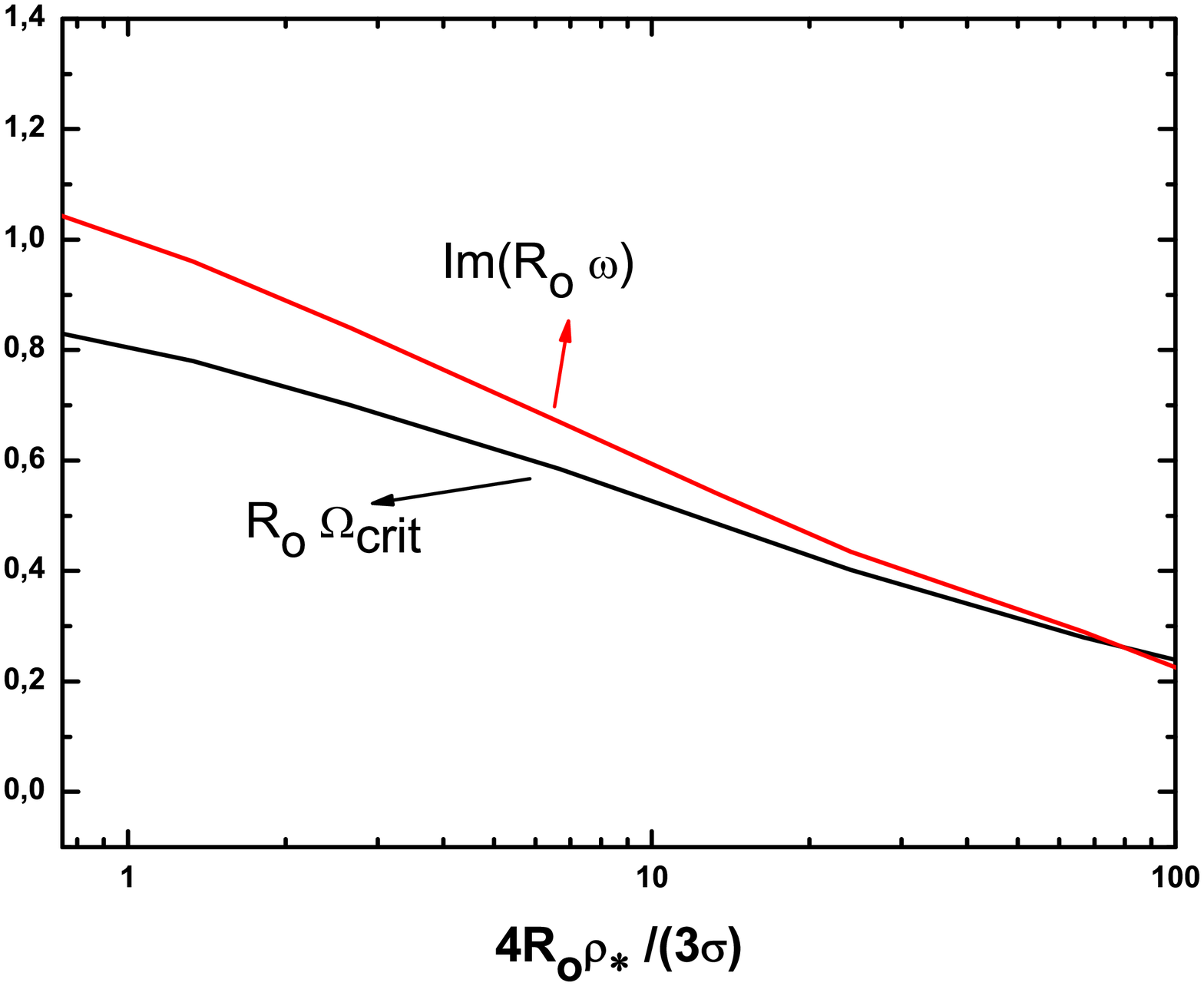,width=8cm,angle=0}
\end{tabular}
\end{center}
\caption{Critical rotation frequency $\Omega R_o$ at which
instability sets  in, as a function of the dimensionless quantity
$\frac{4\rho_{*}R_o}{3\sigma}$. For large
$\frac{4\rho_{*}R_o}{3\sigma}$, the classical prediction
(\ref{classical}) applies.} \label{fig:bifurcation}
\end{figure}

The critical rotation  frequency $R_o\Omega_{\rm crit}$, for which
the configuration is marginally stable is shown in Figure
\ref{fig:bifurcation}. For a given $\rho_{*}R_o/\sigma$ and
rotations larger than $R_o\Omega_{\rm crit}$, the system is unstable
to two lobed perturbations ($m=2$). We find that for moderately
large $\rho_{*}R_o/\sigma\geq 50$, the classical formula
(\ref{classical}) holds.

At this point we should check if and when the lengthscale of the
instability falls within the hydrodynamic limit (discussed in the
end of Section \ref{sec:ballsRings}). The thermodynamic quantities
of the fluid must vary slowly over the mean free path of our system,
$\ell_{\rm mfp}\sim  T_c  \sim \frac{\rho_0}{\sigma}$. The
lenghscale of the instability is the ratio of its wavelength to the
tube's radius and is of order $(\omega R_o^2)^{-1}$. Therefore the
hydrodynamic description is valid for instabilities that satisfy
$\omega R_o\gg \frac{\sigma}{\rho_0 c^2 R_o}$. A simple inspection
of Fig. \ref{fig:bifurcation} allows one to conclude that the
conditions for a hydrodynamic description are satisfied in the
non-relativistic limit where $\frac{\rho_* c^2 R_o}{\sigma}\gtrsim
50$. It is certainly satisfied for very large values of
$\frac{\rho_0 c^2 R_o}{\sigma}$. In addition, the initially
unperturbed plasma ball must of course satisfy the conditions
discussed already in the end of Section \ref{sec:ballsRings}.

\subsection{\label{sec:viscositystab}Viscosity: instability and critical rotation in the  non-relativistic case}

Although we start with a plasma ball in thermodynamic equilibrium, viscosity
contributions have to be taken into account in the perturbed configuration. In fact even the most
``perfect'' known fluids have a certain non-vanishing viscosity
\cite{kovtun}, and this is certainly the case for the SS plasma. It turns out that, for the problem at hand,
even a vanishingly small viscosity has a dramatic effect on the critical rotation at
which instability sets in. In fact, viscosity introduces a
singular-limit, where the ``limit of theory is not the theory of the
limit'' (see \cite{ViscClassic,Scriven1991} for the classical
analysis). Thus our relativistic analysis of the previous subsection
must address also viscosity effects.

Fortunately, from the full relativistic analysis of the previous
section, we know that we can to a good approximation address the
present problem in the small rotation regime. To
define precisely this regime, recall that in section
~\ref{sec:HydroRegime} we concluded that hydrodynamics provides a
good effective description of the deconfined plasma phase immersed
in the vacuum confined phase when the plasma satisfies the condition
\be \frac{\sigma}{\rho_0c^2 R_o}\ll 1\,. \label{HydroReg2}\ee
Moreover, in the previous subsection we found that the full
(numerical) results for the marginally stable mode (where the
$m$-lobed instability set in) agree very well with the (analytical)
non-relativistic results obtained for
\be \frac{\Omega R_o}{c}\ll 1\,. \label{NRlim}\ee
We are therefore justified to use the small velocity regime: in this case not
only the non-relativistic results reveal very good agreement with
the full relativist analysis but also this is the relevant regime
where the hydrodynamic analysis provides valuable information for
the dual gravitational system.

The non-relativistic limit of the hydrodynamic equations presented
in subsection \ref{sec:HydroEqs} was studied in great detail in
\cite{ClassicHydro}. There it is found that the continuity and
Navier-Stokes equations reduce, respectively, to
\begin{eqnarray}
&& \nabla \cdot {\bf v}=0\,,
\nonumber\\
&& \partial_t {\bf v} +({\bf v}\cdot\nabla) {\bf v}=  -
\frac{3}{4\rho_*} \nabla P +\nu \nabla^2 {\bf v} \,, \qquad
\nu\equiv \frac{3\eta}{4\rho_*}\,,
 \label{NR:NavierS}
 \end{eqnarray}
where one uses  $u\rightarrow (1,{\bf v})$, and $\nabla_i$
($i=r,\phi$) represents the covariant derivative with respect to the
purely spatial metric $\eta_{ij}$. So in the non-relativistic limit,
the hydrodynamic system reduces to the continuity and Navier-Stokes
equations for an incompressible fluid with constant density
$\rho{\bigl |}_{\gamma=1}$ (so the continuity equation simply states
that the velocity field is a solenoid vector: its spatial divergence
vanishes), and with kinematical viscosity $\nu$
\footnote{\label{foot}It follows from \eqref{Ball:sol} and
\eqref{TildeRHO:rho0} that in the non-relativistic limit one has
$\rho \sim \frac{4}{3}\rho_*\sim 4\rho_0$. Moreover, in this limit one finds that a possible
contribution coming from the bulk viscosity in the Navier-Stokes
vanishes because it is proportional to $\nabla \cdot {\bf v}$, and
the absence of the particle number conservation and the use of the
Landau frame \cite{Bhattacharyya:2008jc} allow to avoid the use of
thermal conductivity \cite{ClassicHydro}.}.

\begin{figure}[h!tbp]
\begin{center}
\begin{tabular}{l}
\epsfig{file=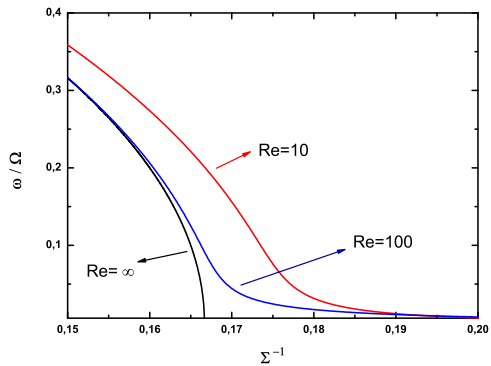,width=8cm,angle=0}
\end{tabular}
\end{center}
\caption{Instability as a function of the inverse of the Bond
number, $\Sigma^{-1}$ for different values of the Reynolds number
$Re=10,\,100$ and for $m=2$. We also show the inviscid values.
Notice that the critical Bond number is {\it not} $m(m+1)$ when
viscosity is present, even in the limit where it vanishes. In the
general viscous case, the instability is limited by the critical
point $\Sigma_c=m^2-1$.} \label{fig:instviscous}
\end{figure}

At this point we can now perturb \eqref{NR:NavierS} and simply
follow the classical analysis of the $m$-lobed instability of a
fluid ball done \eg in \cite{ViscClassic}. Using the same
non-relativistic version of the perturbations
used in \eqref{perturbation}, we get,
\beq && \frac{d}{dr}(r\,U_r)+im U_\phi=0 \,,\nonumber\\
& &  \omega  U_r - 2 \Omega U_\phi+\frac{3}{4 \rho_{*} } {\cal P}'
=\nu \left (U_r''+\frac{1}{r}U_r'-\frac{(m^2+1)}{r^2}U_r-\frac{2im}{r^2}U_{\phi} \right ) \,,\nonumber\\
& & \frac{3}{4\rho_{*}}\frac{im}{r} \calP +2\Omega  U_r+\omega
\, U_\phi =\nu \left
(U_{\phi}''+\frac{1}{r}U_{\phi}'-\frac{(m^2+1)}{r^2}U_{\phi}+\frac{2im}{r^2}U_r \right
)\,.\label{visc}\eeq
These equations must now be supplemented by the appropriate boundary
conditions. The constraint \eqref{constraintYL} is still valid as well
as the associated boundary condition (\ref{3d:bc2}). The Young-Laplace equation gets now a
contribution from the viscous term, and so the normal stress-balance
at the boundary, (\ref{3d:bc1End}), is modified to
\be {\cal P}{\bigl |}_{R_{o}}\simeq \frac{\sigma} {R_{o}} \, \chi_{o}
\lpp \lp  m^2-1\rp -\frac{4\rho_{*}}{3}\,\frac{\Omega^2R_{o}^3}{\sigma}\,\rpp+2\nu \,U_r'\,.
\ee
Finally, we must also require that the tangential stresses vanish at
the boundary (this amounts to require that the fluid is shearless at
the boundary), which yields the extra boundary condition
\be R_{o}^2 U_{\phi}'-R_{o}U_{\phi}-imU_r=0\,. \ee
We can use the first relation in (\ref{visc}) to express $U_\phi$ in
terms of $U_r$ and its derivative and then use the remaining
equations to solve for $U_r$ and ${\cal P}$. One gets a fourth-order
differential equation. In any case, the procedure is trivial and one
ends up with the following eigenvalue equation \cite{ViscClassic}
\be \beta^4+2\beta^2\left ((m^2-1) -m(m-1)^2\frac{\beta^2}{\Phi}
-iRe\right)+m\, Re^2\left (\frac{m^2-1}{\Sigma}-1\right )=0\,,
\label{eigenvalue} \ee
where we defined
\be Re \equiv \frac{R_{o}^2\Omega}{\nu}\,, \qquad \beta^2 \equiv
\frac{\omega}{\Omega} Re\,, \qquad \Phi \equiv
\beta^2+2m-2\beta\frac{I_{m-1}(\beta)+I_{m+1}(\beta)}{2I_{m}(\beta)}
\,. \ee
Here $I_m(\beta)$ is a modified Bessel function and $Re$ is  the
Reynolds number. Solving for this eigenvalue equation, we get the
behavior depicted in Fig. \ref{fig:instviscous}.
What the figure shows, and can be proven be proven analytically
\cite{ViscClassic}, is that the limit of small viscosity,
$Re\rightarrow \infty$,  does {\it not} yield the zero
viscosity result. This is quite an astonishing result: 
the critical Bond number $\Sigma_c$ when viscosity is introduced is
at $\Sigma_c=m^2-1$ that is always smaller than the critical value
for the instability found in the last subsection, namely, $m(m+1)$.
Therefore, the configuration is unstable at {\it lower} rotation
frequencies, when viscosity is accounted for. This result holds for
both large and small Reynolds number.

\begin{figure}[h]
\begin{center}
\small{(a)}\includegraphics[width=6cm]{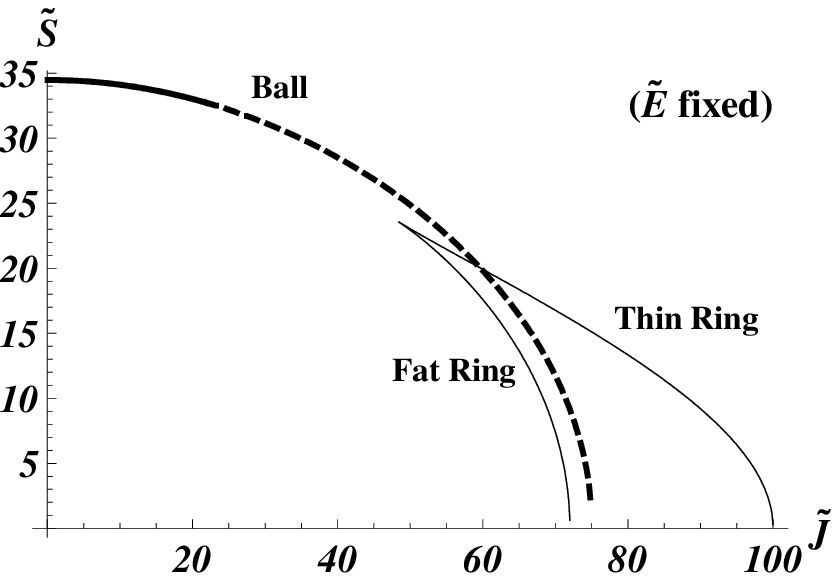}
\hspace{0.5cm}
 \small{(b)} \includegraphics[width=6cm]{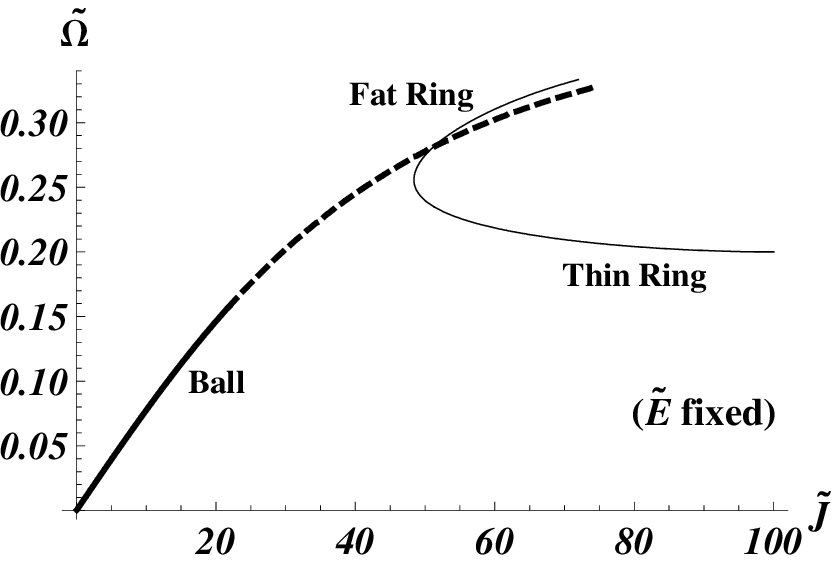}
\caption{(a) Phase diagram with the entropy of plasma balls and
plasma rings $\tE$ as a function of angular momentum $\tJ$, at fixed
energy $\tE=40$ in $d=3$ \cite{Lahiri:2007ae}. For this value of the
energy, one has $\vo^*\simeq 0.7248999$ and $\vi^*\simeq 0.5009615$.
Plasma rings exist only for $\vo \geq \vo^*$. The fat plasma rings
exist for $\tO(\vo=1) \leq\vi \leq\vi^*$ (with
$\tO(\vo=1)=0.333333$), while the thin plasma rings exist for $\vi^*
\leq\vi \leq 1$. At $\vi=\vi^*$ the two families meet in a regular
solution. Hydrodynamics does not provide a good dual description for
configurations near extremality ($T=0$) where the entropy also
vanishes. We find that plasma balls become unstable above the
critical rotation $\tJ_c\simeq 21.8127$, and the dashed line
 represents the unstable plasma balls.
 (b) Similar to Fig a), but this time we represent the phase diagram with
the angular velocity $\tO$ of plasma balls and plasma rings as a
function of angular momentum $\tJ$, at fixed energy, $\tE=40$.
}\label{Fig:PhDiagBallRing:Stab}
 \end{center}
\end{figure}

In Fig.~\ref{Fig:PhDiagBallRing:Stab} we represent the plasma rings,
and stable and unstable plasma balls in two distinct phase diagrams
at fixed energy\footnote{\label{error}In these diagrams we choose to
fix the dimensionless energy at the value $\widetilde{E}=40$ to make
a connection with the value chosen in \cite{Lahiri:2007ae}, where
plasma balls and rings where first discussed. This value of the
energy corresponds to $\frac{\rho_* c^2 R_o}{\sigma}\sim 14$. In
this case the non-relativistic analysis for the marginal point
mismatches the accurate full relativistic result by a factor of
approximately  $20\%$.}, and it summarizes one of our main results.
%

 \setcounter{equation}{0}
\section{\label{sec:bifphase}Bifurcation to two-lobed configurations: rotating plasma peanuts}

In the previous section we found that plasma balls are marginally
stable at $\Sigma_c=m^2-1$, when perturbed by an azimuthal mode $m$.
Such a marginal mode usually signals a branching off to another
family of solutions, and here it is no exception. This new family is
a {\it non}-axisymmetric $m-$lobed configuration: a rotating {\it
plasma peanut}. In this section we want to verify the existence of
this branch of solutions and study some of its main properties. A
detailed numerical study of the full branch of this new plasma
phase is outside of the main scope of this work. Here, we will address the most
important region, namely we investigate how this family looks
close to the bifurcation point $\Sigma=\Sigma_c$ in the phase
diagram of stationary solutions.
Consistent with the analysis of previous sections we restrict
our analysis to the hydrodynamic and small rotation regime, Eqs.
\eqref{HydroReg2} and \eqref{NRlim}, where the hydrodynamic analysis
provides valuable information for the dual gravitational system. We
can then follow closely the analysis done by Benner, Basaran and
Scriven \cite{Scriven1991}. In the following we obtain the new branch
of rotating plasma peanuts that emerges from the plasma ball
bifurcation point.

The energy, angular momentum and entropy of the plasma lump are
given by (\ref{Var:charges}) and (\ref{Var:Entropy}). In the regime
(\ref{HydroReg2}) and (\ref{NRlim}) (see also footnote \ref{foot}
and note that $\rho c^2 +P \sim 4\rho_0 c^2$) and in the close
vicinity of the plasma ball bifurcation point the associated
dimensionless charges read,
\begin{eqnarray}
\widetilde{E}&=&\frac{\rho_0 c^2}{\pi \sigma^2}\,\int_V \,T^{tt}
   \, \approx\, \frac{4}{\pi}\int_0^\pi d\phi\, \widetilde{R}^2(\epsilon,\phi)\,,\nonumber\\
\widetilde{S}&=&\frac{(\rho_0 c^2)^{5/4}}{\pi \alpha^{1/4} \sigma^2}\, \int_V\, \gamma s
 \,\approx\, \frac{4}{\pi}\int_0^\pi d\phi\, \widetilde{R}^2(\epsilon,\phi)\,,\nonumber\\
\widetilde{J}&=&\frac{\rho_0^2 c^5}{\pi \sigma^3}\,\int_V r^2
T^{t\phi}\,\approx\, \frac{2}{\pi}\int_0^\pi d\phi\,
\widetilde{\Omega}(\epsilon) \,\widetilde{R}^2(\epsilon,\phi)  \,.
\label{Var:chargespeanut}
\end{eqnarray}
In these relations we are looking for the unknown family of lobed
solutions by expanding around the known plasma ball at the marginal
stability point $\Sigma_c=m^2-1$. We take $\epsilon$ to be the
parameter that measures the deviation from the rotating axisymmetric
plasma ball with dimensionless radius $\widetilde{R}_o$, and
$\widetilde{R}(\epsilon,\phi)$ describes the non-axisymmetric
boundary of the unknown plasma peanut (note that in the vicinity of
the bifurcation point the axisymmetric deviation is really small and
the perturbative approach is appropriate). We also assume that there
is mirror symmetry around the $\phi=0,\pi$ axis.

In a perturbative analysis, known as power series method
\cite{Scriven1991}, we now take the boundary radius and Bond
parameter to be described by the expansion around the marginally
stable plasma ball,
\beq \widetilde{R}(\epsilon,\phi)&=& \widetilde{R}_0
\left (1+\epsilon f^{(1)}(\phi)+\frac{\epsilon^2}{2} f^{(2)}(\phi) \right )+\mathcal{O}(\epsilon^3)\,,\nonumber\\
\Sigma(\epsilon) &=& \Sigma_c+\epsilon
\Sigma^{(1)}+\frac{\epsilon^2}{2}\,\Sigma^{(2)}+\mathcal{O}(\epsilon^3)\,.
\label{expand1} \eeq
We also introduce the dimensionless pressure jump $\Pi$ at the axis
of rotation (see \eqref{DpEquil}), and its expansion,
\be \Pi \equiv \frac{R_o c^2}{\sigma} \lp
\frac{1}{3}\,\rho_*-\rho_0\rp \,,\qquad \Pi(\epsilon)= \Pi_c
+\epsilon
\Pi^{(1)}+\frac{\epsilon^2}{2}\,\Pi^{(2)}+\mathcal{O}(\epsilon^3)\,.\label{expand2}
\ee
The perturbed quantities characterize the $m$-lobed branch. We want
to find a family of configurations that have fixed energy.
Perturbations have to solve a total of four equations, namely: the
Young-Laplace equation, the energy constraint (we want to represent
the branch of solutions in a phase diagram at fixed energy), the
orthogonality equations and a equation defining the amplitude
parameter $\epsilon$.

\begin{figure}[h]
\begin{center}
\includegraphics[width=7cm]{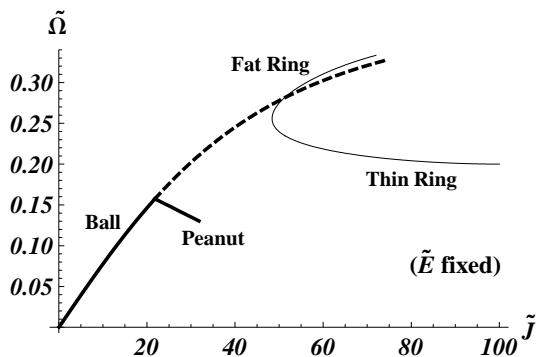} 
\caption{Phase diagram $\tO(\tJ)$ of stationary plasma solutions with fixed energy, $\tE=40$.
At $(\tJ_c,\tO_c)\simeq (21.81,0.157)$ the plasma ball becomes
unstable and bifurcates to a branch of rotating plasma peanuts. We
show this branch in the vicinity of the bifurcation point but
it continues for larger values of $\tJ$. Note that the bifurcation point for plasma balls
is at lower angular velocity and angular momentum than the merging point between the fat
and thin plasma rings, a feature we verified to be valid independently of the
choice for the energy of the system.}\label{Fig:PhDiagPeanut}
 \end{center}
\end{figure}

The perturbed $n^{\rm th}$ order Young-Laplace equation, $\delta
P_<=\sigma \delta K$, yields
\be  f^{(n)}_{\phi\phi}+\lp 1+\Sigma_c \rp
f^{(n)}+\Pi^{(n)}=S_{YL}^{(n)} \,,  \label{expand:YL} \ee
where $f_\phi\equiv \partial_\phi f$ and the relevant source terms
$S_{YL}^{n}$ are
\beq
 S_{YL}^{(0)}&=&0\,,\qquad  S_{YL}^{(1)}= 0\,, \qquad S_{YL}^{(2)}=
(2-\Sigma_c)\lp f^{(1)}\rp^2-2\Sigma^{(1)}f^{(1)}+\lp
f^{(1)}_\phi\rp^2 +
4f^{(1)}f_{\phi\phi}^{(1)} \,, \nonumber \\
S_{YL}^{(3)}&=&-3\Sigma^{(2)}f^{(1)}+3\lp 2-\Sigma_c\rp
f^{(1)}f^{(2)}+3\lp f^{(1)}_\phi
f^{(2)}_\phi-6(f^{(1)})^2f^{(1)}_{\phi\phi}\rp \nonumber \\
 &&  +6\lp
f^{(1)} f^{(2)}_{\phi\phi}+f^{(2)} f^{(1)}_{\phi\phi}\rp
 +9f^{(1)}_{\phi\phi}(f^{(1)}_{\phi})^2-6(f^{(1)}_{\phi})^3-9f^{(1)}(f^{(1)}_{\phi})^2\,.
 \label{expand:YLaux} \eeq
The condition that fixes the energy follows from perturbation of the
first relation in \eqref{Var:chargespeanut}, $\delta
\widetilde{E}=0$, which yields
\be  \int_0^\pi d\phi\,f^{(n)}=S_E^{(n)} \,, \label{expand:E} \ee
with source terms
\be
 S_{E}^{(0)}=0\,,\qquad  S_{E}^{(1)}= 0\,, \qquad S_{E}^{(2)}=
 -\int_0^\pi d\phi\,\lp f^{(1)}\rp^2 \,,
   \qquad S_{E}^{(3)}= 3 \int_0^\pi d\phi\, f^{(1)}f^{(2)} \,.
 \label{expand:Eaux} \ee
For $n\geq 2$ the problem is inhomogeneous and the solution must
satisfy an orthogonality condition\footnote{To get
\eqref{expand:ortho} start with the Young-Laplace equation
\eqref{expand:YL}. Multiply it by $f^{(1)}(\phi)$; integrate over
$\phi$; do a double integration by parts (use the symmetry condition
$f^{(n)}(0)=f^{(n)}(\pi)$ and the $n=1$ Young-Laplace to simplify
some of the terms); and finally make use of the energy conservation
\eqref{expand:E}.},
\be  \int_0^\pi d\phi\,f^{(1)}S^{(n)}_{YL}+2\Pi^{(1)}S_E^{(n)}=0 \,.
\label{expand:ortho} \ee
Finally, the amplitude parameter $\epsilon$ is defined as the
integral-weighted difference between plasma shapes,
\be \epsilon\equiv  \int_0^\pi d\phi\,f^{(1)}\lp
\widetilde{R}-\widetilde{R}_o \rp \quad \rightarrow \quad \int_0^\pi
d\phi\,f^{(1)}f^{(n)}=\delta^1_{n}\,. \label{expand:epsilon} \ee
The solution of \eqref{expand:YL}-\eqref{expand:epsilon},  up to
second order in the perturbation, is
\beq
 f^{(1)}(\phi)&=&\sqrt{\frac{2}{\pi}}\,\cos(m\phi)\,, \nonumber\\
f^{(2)}(\phi) &=&\frac{3}{\pi}\left
(1-\frac{1}{m^2}\right)+\frac{2}{\pi}\left
(\frac{1}{m^2}-1\right)\cos^2(m\phi)
+\frac{2}{\pi}\left (\frac{2}{m^2}-3\right)\sin^2(m\phi)\,,\nonumber\\
\hspace{-0.5cm}\Sigma &=&
(m^2-1)+\frac{\epsilon^2}{2}\,\frac{3}{2\pi}\,\frac{(m^4-1)(1-m^2)}{m^2}\,,
\nonumber\\
\hspace{-0.5cm}\Pi &=&
1-\frac{m^2-1}{2}+\frac{\epsilon^2}{2}\,\frac{3}{\pi}\,(1-m^2)\,.
\eeq
These perturbations keep the energy of the stationary solutions
fixed and equal to $\tE = 4 \widetilde{R}_o^2$. Moreover, use of
these relations in the definition (\ref{Bond}) of the Bond
parameter and in (\ref{Var:chargespeanut}) yields the expansion of
the other thermodynamic quantities around the bifurcation point,
\beq
\tO &=&\widetilde{R}_o^{-3/2} \sqrt{m^2-1}\left (\frac{1}{2}-\epsilon^2\,\frac{3(m^4-1)}{16m^2\pi}\right )
+{\cal O} \left (\epsilon^3\right )\,,\nonumber\\
\tJ&=&\widetilde{R}_o^{5/2}\sqrt{(m^2-1)}\left
(1+\epsilon^2\,\frac{3+32m^2-3m^4}{8\pi\,m^2}\right)
 +{\cal O} \left (\epsilon^3\right )\,,\nonumber\\
\tS&=&4\widetilde{R}_o^2 +{\cal O} \left (\epsilon^3\right )\,. \eeq
This expansion allows to represent the new branch of
non-axisymmetric plasma peanuts in a phase diagram of stationary
solutions at fixed energy. Since rotation is the mechanism
responsible for the instability that signals the bifurcation to the
new phase, it is appropriate to represent the 2-lobed plasma lump in
a phase diagram that represents the angular momentum $\widetilde{J}$
against its conjugated chemical potential $\widetilde{\Omega}$. To
check the accuracy of our approximations we first observe that at
zero order, we have
$\tJ^{(0)}=2\widetilde{R}_o^4\tO=\frac{2}{16}\tE^2 \tO$. For the
slope of the plasma ball at the bifurcation point we thus get
$\frac{d\tO^{(0)}}{d\tJ^{(0)}}=\frac{1}{200}$ for $\tE=40$. This
value is in reasonable agreement with the exact relativistic value
for the $\widetilde{\Omega}-\widetilde{J}$ slope at the bifurcation,
obtained from \eqref{Ball:charges} (for $\widetilde{E}=40$ we get
numerically $\frac{d\tO^{(0)}}{d\tJ^{(0)}}\sim \frac{1.2}{200}$; as
explained in footnote \ref{error} the disagreement is due to the
fact that the system with $E=40$ is not exactly within the classical
regime). We can now use the next-to-leading order non-vanishing
contribution to get the desired slope for the $m$-lobed branch,
yielding
\beq \left (\frac{d\tO}{d\tJ}\right )
^{(2)}&=&-\frac{24(m^4-1)}{3+32m^2-3m^4}\frac{1}{\tE^2}\,.
\label{slope:peanut}\eeq
For $m=2$ this is the slope at which the 2-lobed or plasma peanut
branch emerges from the plasma ball bifurcation point. In the phase
diagram $\widetilde{\Omega}-\widetilde{J}$ of stationary solutions
at fixed energy, represented in Fig. \ref{Fig:PhDiagPeanut}, the
plasma peanut branch bends down and to the right relatively to the
plasma ball family. In this diagram we identify the critical
unstable point where bifurcation occurs and we represent by a dashed
line the plasma balls rotating faster than the critical velocity,
which are therefore unstable.

This figure summarizes the main result of this section: we have
confirmed the existence of a new phase of non-axisymmetric
stationary solutions and we were able to use perturbative methods
around the bifurcation point to find what is the direction that the
new branch of solutions takes relatively to the known phases. A full
description of the 2-lobed branch well away from the bifurcation
point (where it acquires a well defined peanut shape) would required
a full numerical analysis and we leave it for future work. Note also
that keeping going up along the plasma ball branch this time already
in the unstable ball region to 2-lobed perturbations we would find a
succession of new bifurcation points to new phases of solutions
representing $m$-lobed plasma lumps with $m>2$. The slope of these
branches is given by (\ref{slope:peanut}).

Alternatively, we could represent the new phase in the $\tS-\tJ$
phase diagram. We get $(d\tS/d\tJ)^{(2)}=0$ up to second order terms
(this is not a surprise since at the order we work, the entropy is
equal to the energy). Now, in the classical regime, the plasma
family also has zero slope for this quantity so in principle one
should go to higher order in $\epsilon$ if we wish to compute
accurately the slope of the $m-$lobed family. In our approximation
the bifurcation point tends to collapse to the static plasma ball
point in the $\tS-\tJ$ diagram where the slope vanishes. Therefore,
at the order we work, the representation in this diagram is not a
good choice.

The current line of research is not exhausted in the present
analysis. In future work, it should be certainly possible to find
numerically the full branch of non-axisymmetric plasma lumps in the
phase diagram, and to discuss their stability.

 \setcounter{equation}{0}
\section{Unstable plasma balls and their dual black holes\label{sec:BallBH}}

In the previous two sections we found that, starting with a static
plasma ball, as we increase its rotation, the axisymmetric rotating
plasma ball becomes unstable first to a $2$-lobed perturbation and
then to $m>2$ lobed perturbations. Moreover, the marginal unstable
modes were found to be bifurcation points in the phase diagram of
solutions to a new branch of $m$-lobed plasma lumps. It is important
to emphasize that these new plasma phases inhabit a region in
parameter space where hydrodynamics provides a reliable holographic
description of the dual gravitational system. This dual system was
discussed in the Introduction: in the long wavelength regime, $3d$
fluid dynamics is an effective theory describing the Scherk-Schwarz
(SS) compactification of a $4d$ CFT. The latter is dual to the SS
compactification of AdS$_5$ gravity. It then follows that $3d$
plasma lumps correspond, in the dual gravity description, to SS
AdS$_5$ black objects. The axisymetric plasma balls and plasma rings
correspond, respectively, to rotating black holes and black rings in
SS AdS$_5$ \cite{Lahiri:2007ae}.

Since plasma balls are unstable above a critical rotation rate (cf.
Section~\ref{sec:stability}), the holographic dual SS AdS$_5$ black
holes must also be unstable against $m$-lobed perturbations. These
axisymmetric black holes are expected to bifurcate to a new branch
of {\it non}-axisymmetric black hole solutions in accordance with
the plasma results. In the simplest $m=2$ case, these are
non-axisymmetric solutions that describe a peanut-like rigidly
rotating black hole. Such a deformed black hole must necessarily
emit all kinds of radiation. These waves can escape to infinity
through the directions parallel to the holographic boundary (where
the dual fluid lives). It is then natural to expect that this
non-axisymmetric black hole will decay, probably emitting a good
amount of the angular momentum and multipoles of the system, into an
axisymmetric slowly rotating black hole.\footnote{We are undoubtedly
grateful to Roberto Emparan for fruitful discussions concerning the
dual interpretation of the plasma results} In practice, this
instability provides then a mechanism that constraints the rotation
of the black hole: effectively it introduces an upper bound for the
rotation of the SS black hole. Some of these non-axisymmetric black
holes are expected to decay very slowly and to be long-lived. Indeed
it is important to emphasize that hydrodynamics can only provide a
good description of a gravitational system in a regime where the
gravitational interaction and radiation is suppressed. For example,
two plasma balls do not interact and their collision is not
accompanied by radiation emission, so they can only describe
approximately two black holes when they are widely separated.
Therefore, in the regime where our fluid description provides a good
approximation, we should expect radiation emission to be suppressed
in the dual gravitational system. Thus the non-axisymmetric black
holes should indeed be long-lived, leaking very slowly radiation and
angular momentum. Finding the explicit geometry describing such a
solution will probably require a full-blown numerical solution of
the field equations. This is thus a valuable example of the power of
the hydrodynamic/gravity duality.

We expect our main conclusions to extend to higher dimensional
theories as well. For any $d\geq 3$, $d$-dimensional fluid dynamics
is an effective theory describing the SS compactification of a
$(d+1)$-dimensional CFT, which is dual to a SS compactification of
AdS$_{d+2}$. Apart from an expected simplification of the technical
analysis in the $d=3$ case, there is clearly no step in our analysis
that is valid only for $d=3$. Thus, plasma balls and dual black
holes rotating above a critical rate should also be unstable in
higher dimensions and bifurcate to $m$-lobed plasmas and black
holes. But, as we discuss in the sequel, we also expect some
significant differences between the $d=3$ and $d>3$ cases.

Eventually new features are expected as one climbs the
dimensionality ladder. For instance, new plasma lump configurations
appear for $d>3$. Indeed, in $d>3$ one can have, besides the
configurations discussed here, also pinched plasma balls,
non-uniform tubes and pinched non-uniform tubes
\cite{Lahiri:2007ae,Bhardwaj:2008if,Caldarelli:2008mv}. In the dual
gravitational system, pinched balls \cite{Bhardwaj:2008if}
correspond to pinched black holes \cite{Emparan:2003sy}, and the
non-uniform tubes describe non-uniform black strings with the
pinched ones being black strings on the verge of expelling a black
ring \cite{Caldarelli:2008mv}. It would be quite interesting to
understand the stability properties and phase diagrams of plasma
balls in higher dimensions.

But, the results of the present analysis already allow one to infer
important properties about the stability and existence of new
solutions in higher dimensions. Consider hydrodynamics in a $d=4$
Minkowski background (the boundary of SS AdS$_6$ black objects).
Take a static plasma tube with topology $D^2\times \mathbb{R}$ or
$D^2\times S^1$ (it is translationally invariant along the extra
direction which can be or not compact). Such a plasma tube is
unstable against the Rayleigh-Plateau instability when its length is
larger than its transverse perimeter
\cite{Cardoso:2006ks,Cardoso:2006sj,Caldarelli:2008mv}. In the dual
system, the corresponding black string is Gregory-Laflamme unstable
\cite{Gregory:1993vy}, the holographic dual of the fluid
instability. The marginal unstable mode is a bifurcation point to a
new phase of solutions describing non-uniform plasma tubes and black
strings. For details of this instability, see Refs.
\cite{Cardoso:2006ks,Cardoso:2006sj,Caldarelli:2008ze,Caldarelli:2008mv,mandm}.
Now, a $4d$ static plasma tube is simply a $3d$ plasma ball
trivially extended along the extra direction. Our results then show
that a rotating plasma tube, with the rotation axis along the tube
direction, should also be unstable, above a critical rotation rate,
to $m-$lobed perturbations. Moreover, the marginally stable points
are bifurcation points, this time to non-axisymmetric plasma tubes
translationally invariant along the tube direction and whose
transverse cross section has a $m$-lobed shape. Again, by the
hydrodynamic/gravity duality, we expect that rotating black strings
will become unstable, not only against the Gregory-Laflamme
instability, but also against $m$-lobed azimuthal perturbations. And
a new branch of non-axisymmetric $m$-lobed black strings is expected
to branch-off at the unstable threshold point in the phase diagram
of solutions. Other interesting solutions include axisymmetric
non-uniform black strings (these bifurcate from the Gregory-Laflamme
unstable point) and pinched non-uniform black strings which are also
solutions of the gravity theory \cite{Caldarelli:2008mv}. Our study
then predicts the existence of non-axisymmetric non-uniform black
strings and pinched ones. We emphasize that the existence of pinched
black strings \cite{Caldarelli:2008mv} and non-axisymmetric
$m$-lobed black strings would hardly be anticipated without
resorting to the hydrodynamic/gravity duality. Our discussion
focused on $4d$ plasma tubes dual to $6d$ black strings but again it
should extend to higher dimensions.

These results rely on the hydrodynamic description of a particular
gauge/gravity duality, namely of the Scherk-Schwarz system.
Experience with hydrodynamics indicates that similar results should
be found for other gauge/gravity theories with an effective
hydrodynamic description and a confinement/deconfinement phase
transition. The deconfined black hole phase of such a general system
is still expected to be described at leading order by a perfect
fluid holographic stress tensor, and the interface between the
deconfined and confined phases is again expected to be dictated by a
domain wall with a surface tension. Such a generic dual system will
of course have a different equation of state, encoding the
information on the kind of fluid describing the gauge theory.
However, our analysis is not very sensitive to the particular
equation of state of the plasma. Hence it could be that our main
results on the stability and bifurcation properties of plasma balls
will be common to generic dualities.

It would also be interesting to investigate the dual of the
superradiant instability on the plasma balls. In general, rotating
black holes develop an ergoregion, a region in spacetime with
negative energy states. For spacetimes with ergoregions, one can
have superradiant scattering, whereby a wave (with frequency
$\omega<m\Omega$, where $m$ is an azimuthal number and $\Omega$ is
the horizon velocity) can be amplified, extracting rotational energy
from the hole. In AdS, the superradiantly amplified waves bounce
back at infinity and lead to a superradiant instability
\cite{Cardoso:2004nk}. Thus four-
\cite{Cardoso:2006wa,Cardoso:2004hs} and higher dimensional
\cite{Kunduri:2006qa,superradiance} rotating Kerr-AdS black holes
can be unstable. The endpoint of such an instability is presumably
an element of a new branch of stationary black holes, rotating in
such a way as to avoid the superradiant window
\cite{Kunduri:2006qa,Cardoso:2006wa}. Some of the angular momentum
of the black hole is transferred to caged radiation in between the
horizon and the AdS wall and co-rotating with the black hole. It was
also argued in \cite{Kunduri:2006qa} that a new branch of
non-axisymmetric black hole solutions could eventually bifurcate
from the original Kerr-AdS black hole at the threshold of the
superradiant instability. This superradiant phenomena is a similar,
but not identical, mechanism to the one we explored in this paper
(the mechanism dealt with here is not dependent on superradiant
amplification). They may correspond to different bifurcation
branches in a phase diagram of possible solutions.
 To explore the eventual dual of the superradiant instability on a
plasma ball we should take into account that experience indicates
that an ergoregion instability develops when the rotation speed at
the boundary surface, $\Omega R_o$, exceeds the sound speed of the
plasma. This then corresponds to the formation of an acoustic
``ergoregion" in the system, and therefore by a general theorem by
Friedman \cite{friedman}, they should be unstable. We should however
keep in mind a possible serious caveat: the dual hydrodynamic
description is typically valid for large AdS black holes (\ie for
those whose horizon is large compared with the cosmological scale)
\cite{Lahiri:2007ae,Bhattacharyya:2007vs}, while the superradiant
instability is present only in small AdS black holes
\cite{Cardoso:2006wa,Cardoso:2004hs}. So it might well be that the
fluid description is not able to capture the dual of the
superradiant instability.

\section{Acknowledgments}
We warmly thank Marco Caldarelli, Roberto Emparan and Dietmar Klemm
for very fruitful discussions, and Roberto Emparan for his useful
comments to the final version of this manuscript. We also thank CERN
for hospitality during the programme ``Black Holes: A Landscape of
Theoretical Physics Problems", August-October 2008, where part of
this work was done. OJCD further acknowledges the University of
Barcelona where this work started and the organizers of the workshop
``Higher dimensional black holes: Exact solutions and their
stability", Laboratoire de Physique Th\'eorique, France. This work
was partially funded by Funda\c c\~ao para a Ci\^encia e Tecnologia
(FCT) - Portugal through projects PTDC/FIS/64175/2006 and
CERN/FP/83508/2008. VC acknowledges financial support through a
Fulbright Scholarship. OJCD acknowledges financial support provided
by the European Community through the Intra-European Marie Curie
contract MEIF-CT-2008.

\appendix

\section*{Appendices}

 \setcounter{equation}{0}
\section{On the stability of plasma rings \label{sec:StabRings}}

In this Appendix we discuss the stability of plasma rings,
and we find that plasma rings are {\it stable} against the {\it
particular} $m$-lobed deformations that we consider.
Plasma rings have an inner boundary at $r=R_{\rm i}$ in addition to
the outer boundary at $r=R_o$. The boundary conditions
(\ref{3d:bc1End}) and (\ref{3d:bc2}) yield the following system for
the outer and inner surfaces,
\be U_r(R_{o,{\rm i}})=
 \frac{\omega R_{o,{\rm i}}^2\,\gamma_{o,{\rm i}}^{8}}{\sigma\,\lpp
 \pm \lp \frac{1}{c^2}(\omega-im\Omega)^2 R_{o,\rm i}^2 + m^2-1\rp
 -\frac{4\rho_{*}}{3}\,\frac{\Omega^2R_{o,\rm i}^3}{\sigma}\,\gamma_{o,\rm i}^6 \rpp }
\,\calP(R_{o,{\rm i}})\,.\label{Ring:secondBC}\ee
Relations (\ref{vasP}) and (\ref{Ring:secondBC}) provide two
conditions on $\calP$ and constitute again an eigenvalue problem for
$\omega$. The problem depends on two dimensionless quantities:  $\frac{\rho_{*} R_o}{\sigma}$ and $\Omega
R_o$. For $\Omega R_o\ll c$, (\ref{ODE}) and
(\ref{vasP}) reduce respectively to (\ref{NR:1}) and (\ref{NR:2}),
while (\ref{Ring:secondBC}) simplifies to
\be
 U_r(R_{o,{\rm i}})=\frac{\omega R_{o,{\rm i}}^2}{\sigma\,\lp
\pm\lp m^2-1\rp-\frac{4\rho_{*}}{3}\frac{\Omega^2 R_{o,{\rm
i}}^3}{\sigma} \rp} \,\calP(R_{o,{\rm i}})\,.
\label{Ring:secondBC:dimless}\ee
For black rings some of these quantities are not independent,
\eg we have $R_{\rm i}=R_{\rm i}(R_o)$ and
$\Omega=\Omega(R_o,\rho/\sigma)$. In section \ref{sec:ballsRings} we
already discussed these constraints so it is appropriate to rewrite
(\ref{NR:1}), (\ref{NR:2}) and (\ref{Ring:secondBC:dimless}) in
terms of the dimensionless variables (\ref{newvars}). Defining also
the dimensionless frequency
$\widetilde{\omega}=\frac{\sigma}{\rho_0}\,\omega$ the system we
have to solve is
\beq
 && {\cal P}(v)=Av^m+B v^{-m}\,, \label{Ring1}\\
 &&
 -\widetilde{\Omega}^2\,\frac{2im\widetilde{\Omega}{\cal P}(\vo)+\widetilde{\omega} \vo {\cal P}'(\vo)}
 {4v g_+(\vo) \lp\widetilde{\omega}^2 +4\widetilde{\Omega}^2\rp }
 = \frac{\widetilde{\omega}}{\widetilde{\Omega}}\,\frac{ \vo^2}
 {\lpp \pm\lp m^2-1\rp-\frac{4}{\widetilde{\Omega}}g_+(\vo)\vo^3\rpp}
\,\calP(\vo) \,,
 \label{Ring2}\\
&&
 -\widetilde{\Omega}^2\,\frac{2im\widetilde{\Omega}{\cal P}(\vi)+\widetilde{\omega} \vi {\cal P}'(\vi)}
 {4v g_-(\vi)  \lp\widetilde{\omega}^2 +4\widetilde{\Omega}^2\rp }
 = \frac{\widetilde{\omega}}{\widetilde{\Omega}}\,\frac{ \vi^2}
 {\lpp \pm\lp m^2-1\rp-\frac{4}{\widetilde{\Omega}}g_-(\vi)\vi^3\rpp}
\,\calP(\vi)\,. \label{Ring3}\eeq

We have found no unstable mode. One might argue that black rings
should be unstable at least at the point where they cross that
plasma ball diagram of Fig.~\ref{Fig:PhDiagPeanut}. This does not
necessarily have to be in contradiction with our results first
because the boundary conditions for black rings do not seem to
reduce to the boundary conditions for plasma balls, even in this
limit. Second, the analysis here does not prove that these plasma
rings are stable: they are stable against the particular $m$-lobed
deformations that we consider (see discussion just after
\eqref{displacementd3}). A more general analysis is needed,
encompassing generic perturbations.


\end{document}